\documentclass[aps,prd,amsmath
,eqsecnum            
,preprintnumbers
,nofootinbib         
 ,showpacs          
 ,showkeys          
,twocolumn         
]{revtex4}
\usepackage[dvips]{graphicx}
\usepackage{amsmath}
\usepackage{amsfonts}
\usepackage{amssymb}
\usepackage{longtable}   

\allowdisplaybreaks[4]     

\newcommand{\df}{\ {\overset {\rm def} =}\ }
\newcommand{\dr}[2]{\frac {{\rm d} {#1}} {{\rm d} {#2}}}

\newcommand{\dril}[2]{{{\rm d} {#1}} / {{\rm d} {#2}}}

\newcommand{\llim}[1] {\ {\underset {#1} {\longrightarrow}}\ }

\begin{document}

\title{Accelerating expansion or inhomogeneity? \\
A comparison of the $\Lambda$CDM and Lema\^{\i}tre -- Tolman models}

\author{Andrzej Krasi\'nski}
\affiliation{N. Copernicus Astronomical Centre, Polish Academy of Sciences, \\
Bartycka 18, 00 716 Warszawa, Poland} \email{akr@camk.edu.pl}

\date {}

\begin{abstract}
It is shown how certain observations interpreted in the background of the
Friedmann model with $\Lambda < 0 = k$ (the $\Lambda$CDM model) can be
re-interpreted using the $\Lambda = 0$ Lema\^{\i}tre -- Tolman (L--T) model so as
to do away with the ``dark energy''. The purpose of the paper is to clarify the
underlying geometrical relations by doing the calculations as much as possible
analytically or by very simple numerical programs. In the first part of the paper
(fictitious) observations of the distribution of expansion velocity along the
past light cone of the observer are considered. It is shown that the whole past
light cone of the $\Lambda$CDM observer can be reproduced in the L--T model with
$\Lambda = 0 = E$. This is a geometric exercise that has the advantage of being
free of numerical complications. In the second part, the luminosity distance --
redshift relation of the $\Lambda$CDM model is duplicated using the L--T model
with $-k = 2E/r^2 =$ constant $> 0$. The value of $k$ and the function $t_B(r)$
are determined by the $\Lambda$CDM parameters. General properties of this L--T
model are described. Difficulties of carrying the numerical calculations through
the apparent horizon are presented in detail and mostly solved. The second model
is a counterexample to the general belief that an L--T model mimicking
$\Lambda$CDM must contain a void around the center -- it has a peak of density at
$R = 0$.
\end{abstract}

\maketitle

\section{Accelerating expansion or inhomogeneity?}

\setcounter{equation}{0}

As is well-known by now, in the years 1998 -- 1999 two teams of observers
\cite{Perl1999,Ries1998} concluded that the observed peak luminosity of the type
Ia supernovae is smaller than was implied by a $\Lambda = 0$ Friedmann model. An
elaborate fitting procedure led to the conclusion that the best-fit model {\em
within the Robertson -- Walker (RW) class with zero pressure} is the one with
the curvature index $k = 0$ and a value of the cosmological constant that
accounts for $\approx 68$\% of the current energy-density of the Universe
\cite{Plan2013}, now called the $\Lambda$CDM model. Thus, at present, the
Universe should be expanding at an accelerating rate. The substance that causes
this acceleration was named ``dark energy''. Strange as it is (an observed
effect being caused by an entity that no-one has ever seen outside this
cosmological context), this hypothesis was almost universally accepted, and the
existence of the dark energy is now taken for granted by nearly all authors.

Meanwhile, it has been demonstrated in several papers that if one gives up on the
homogeneity assumption, then even the simplest among the realistic inhomogeneous
models, the Lema\^{\i}tre \cite{Lema1933} -- Tolman \cite{Tolm1934} (L--T) model,
can account for the apparent dimming of the type Ia supernovae using a suitable
inhomogeneous distribution of mass in the Universe, with zero cosmological
constant and decelerated expansion. Among the first papers that introduced this
alternative description were the ones by Celerier \cite{Cele2000} and by Iguchi,
Nakamura and Nakao \cite{INNa2002}. Later, it was demonstrated by examples that
when the L--T model is employed at full generality, with no a priori simplifying
assumptions, then two sets of observational data can be reproduced, for example
the pairs (angular diameter distance -- mass density in the redshift space) and
(angular diameter distance -- expansion rate) \cite{CBKr2010}.

Those earlier considerations resorted to numerical calculations almost from the
beginning, which obscured the underlying geometrical relations. In the present
paper, a comparison of the $\Lambda$CDM model with the $\Lambda = 0$ L--T model
is done by more transparent means. Explicit algebraic and differential equations
are used almost exclusively, and several properties of the L--T model thus
adapted are determined by exact calculations.

In the first part of the paper (Sections \ref{tilt} -- \ref{comments}) the
distribution of the cosmic expansion velocity along the past light cone of the
observer is considered. It is shown that, with a suitably chosen bang-time
function $t_B(r)$, the central observer in the L--T model with $E = 0 = \Lambda$
can see {\it the same} past light cone as an observer in the $\Lambda$CDM model.
This proof is unrelated to actual problems of observational cosmology, but it is
free of numerical complications, and therefore is presented first.

In the second part (Sections \ref{duplicate} -- \ref{LTcosmology}, inspired by
the approach of Iguchi et al. \cite{INNa2002}), the luminosity distance --
redshift relation, $D_L(z)$, of the $\Lambda$CDM model is duplicated in the L--T
model with $\Lambda = 0$ and $-k = 2E/r^2 =$ constant $> 0$ (this is the same $k$
as in the limiting Friedmann model). The value of $k$ is determined by
fine-tuning the values of redshift at the origin and at the apparent horizon, and
the effect of $\Lambda$ is reproduced by the L--T bang time function $t_B(r)$.

The L--T model mimicking the $\Lambda$CDM $D_L(z)$ relation is determined for a
single instant of observation. The time-evolution of the two models is
different, and they can be distinguished by observations that are sensitive to
time-changes rather than just to an instant snapshot of the Universe, for
example by the redshift drift \cite{QABC2012}.

The approach used here leads to a few clarifications. Among other things, it is
shown how the obstacles to carrying the numerical integration through the
apparent horizon, reported in Refs. \cite{INNa2002,VFWa2006} (and incorrectly
interpreted in \cite{VFWa2006} as a ``pathology'' of the L--T model), can be
overcome. Also, the model considered in the second part provides a counterexample
to the general belief that an L--T model mimicking accelerated expansion must
contain a void around its center of symmetry.

\section{A quick introduction to the Friedmann and Lema\^{\i}tre -- Tolman
models}\label{LTintro}

\setcounter{equation}{0}

This is a summary of basic facts about the L--T model. For extended expositions
see Refs. \cite{Kras1997,PlKr2006}. Its metric is:
\begin{equation}\label{2.1}
{\rm d} s^2 = {\rm d} t^2 - \frac {{R_{,r}}^2}{1 + 2E(r)}{\rm d} r^2 -
R^2(t,r)({\rm d}\vartheta^2 + \sin^2\vartheta \, {\rm d}\varphi^2),
\end{equation}
where $E(r)$ is an arbitrary function, and $R(t, r)$ is determined by the
integral of the Einstein equations:
\begin{equation}\label{2.2}
{R_{,t}}^2 = 2E(r) + 2M(r) / R - \tfrac 1 3 \Lambda R^2,
\end{equation}
$M(r)$ being another arbitrary function and $\Lambda$ being the cosmological
constant. Note that $E$ must obey
\begin{equation}\label{2.3}
2E + 1 \geq 0
\end{equation}
in order that the signature of (\ref{2.1}) is the physical $(+ - - -)$. The
equality in (\ref{2.3}) can occur only at special locations (at isolated values
of $r$) called necks \cite{PlKr2006}.

Equation (\ref{2.2}) has the same algebraic form as one of the Friedmann
equations, except that it contains arbitrary functions of $r$ in place of
arbitrary constants. The solution of (\ref{2.2}) may be written as
\begin{equation}\label{2.4}
t - t_B(r) = \pm \int \frac {{\rm d} R} {\sqrt {2E(r) + 2M(r) / R - \tfrac 1 3
\Lambda R^2}},
\end{equation}
where $t_B(r)$ is one more arbitrary function called the bang time. The $+$ sign
applies for an expanding region, $-$ applies for a collapsing region. Throughout
this paper only expanding models will be considered.

In the case $\Lambda = 0$, the solutions of (\ref{2.2}) may be written in the
parametric form as follows:

(1) When $E(r) < 0$:
\begin{eqnarray}\label{2.5}
R(t,r) &=& - \frac M {2E} (1 - \cos \eta), \nonumber \\
\eta - \sin \eta &=& \frac {(-2E)^{3/2}} M \left[t - t_B(r)\right].
\end{eqnarray}

(2) When $E(r) = 0$:
\begin{equation}\label{2.6}
R(t,r) = \left\{\frac 9 2 M(r) \left[t - t_B(r)\right]^2\right\}^{1/3}.
\end{equation}

(3) When $E(r) > 0$:
\begin{eqnarray}\label{2.7}
R(t,r) &=& \frac M {2E} (\cosh \eta - 1), \nonumber \\
\sinh \eta - \eta &=& \frac {(2E)^{3/2}} M \left[t - t_B(r)\right].
\end{eqnarray}

The mass density is
\begin{equation}  \label{2.8}
\kappa \rho = \frac {2{M_{,r}}}{R^2R_{,r}}, \qquad \kappa \df \frac {8\pi G}
{c^2}.
\end{equation}
The pressure is zero, so the matter (dust) particles move on geodesics.

Equations (\ref{2.1}) -- (\ref{2.8}) are covariant with the transformation $r
\to r' = f(r)$, which may be used to give one of the functions $(M, E, t_B)$ a
handpicked form, in the range where it is monotonic. In this paper, $M,_r > 0$
is assumed, and the following choice of $r$ will be made
\begin{equation}\label{2.9}
M = M_0 r^3,
\end{equation}
where $M_0 > 0$ is an arbitrary constant. This $r$ is still not unique -- the
transformations $r = Cr'$, with $C =$ constant, are still allowed, and they
redefine $M_0$ by $M_0 = M'_0 / C^3$. So, we can assume a convenient value for
$M_0$. However, $M_0$ has the dimension of length and represents mass, so the
choice of its value amounts to choosing a unit of mass. See Sec.
\ref{numerunits} for more on this.

As seen from (\ref{2.8}), the locus of $R,_r = 0$ is a curvature singularity
($\rho \to \infty$), unless it coincides with the locus of $M,_r = 0$ -- but
this last one is absent here because of (\ref{2.9}). This singularity is called
{\em{shell crossing}} because, as seen from (\ref{2.1}), the geodesic distance
between the $r$- and $(r + {\rm d} r)$ spheres becomes zero there. The full set
of necessary and sufficient conditions for avoiding shell crossings was worked
out in Ref. \cite{HeLa1985}. With the assumption $M,_r > 0$, and $E,_r > 0$
adopted further on, the necessary and sufficient condition for the absence of
shell crossings is
\begin{equation}\label{2.10}
\dr {t_B} r < 0.
\end{equation}
In the case $E = 0$, $R,_r = 0$ implies, via (\ref{2.6}) and (\ref{2.9})
\begin{equation}\label{2.11}
t - t_B(r) = \tfrac 2 3 r \dr {t_B} r.
\end{equation}
Since $r > 0$ by assumption (\ref{2.9}), and $t > t_B$ in expanding models,
(\ref{2.11}) has no solutions when $\dril {t_B} r < 0$.

It must be stressed that the L--T model, having zero pressure, cannot be applied
to those cosmological situations, in which pressure cannot be neglected, in
particular to the pre-recombination epoch. Consequently, if a shell crossing
exists, but occurs before last scattering (usually assumed to take place between
$3 \times 10^5$ and $4 \times 10^5$ y after the Big Bang), then it is
cosmologically irrelevant -- the L--T model does not apply to those times
anyway.

A past radial null geodesic is given by the equation
\begin{equation}\label{2.12}
\dr t r = - \frac {R_{,r}} {\sqrt{1 + 2E(r)}},
\end{equation}
and its solution is denoted $t = t_{\rm ng}(r)$. The redshift $z(r)$ along
$t_{\rm ng}(r)$ is given by \cite{Bond1947,PlKr2006}:
\begin{equation}\label{2.13}
\frac 1 {1 + z}\ \dr z r = \left[ \frac {R_{,tr}} {\sqrt{1 + 2E}} \right]_{\rm ng}.
\end{equation}
Given $t_{\rm ng}(r)$ and $z(r)$, the luminosity distance $D_L(z)$ of a light
source from the central observer is \cite{Cele2000,BKHC2010}
\begin{equation}\label{2.14}
D_L(z) = (1 + z)^2 \left.R\right|_{\rm ng}.
\end{equation}

The Friedmann limit of (\ref{2.1}) follows when $M/r^3 = M_0$, $2E / r^2 = - k$
and $t_B$ are constant, where $k$ is the Friedmann curvature index. Then
(\ref{2.5}) -- (\ref{2.7}) imply $R = r S(t)$,\footnote{A coordinate-independent
condition for the Friedmann limit is $2E/M^{2/3}$ and $t_B$ being constant. Then
$R = [M(r)/M_0]^{1/3} S(t)$.} and the limiting metric is
\begin{equation}\label{2.15}
{\rm d} s^2 = {\rm d} t^2 - S^2(t) \left[\frac 1 {1 - kr^2} {\rm d} r^2 +
r^2 ({\rm d}\vartheta^2 + \sin^2\vartheta \, {\rm d}\varphi^2)\right].
\end{equation}
Equation (\ref{2.13}), using (\ref{2.12}), simplifies to $(\dril z t)/(1 + z) =
S,_t/S$, which is easily integrated to give
\begin{equation}\label{2.16}
1 + z = S(t_o)/S(t_e),
\end{equation}
where $t_o$ and $t_e$ are the instants of, respectively, observation and emission
of the light ray.

In the Friedmann limit, the formula for the luminosity distance can be
represented as follows
\begin{eqnarray}\label{2.17}
&& D_L(z) = \frac {1 + z} {H_0 \sqrt{\Omega_k}} \\
&& \times \sinh \left\{\int_0^z \frac
{\sqrt{\Omega_k} {\rm d} z'} {\sqrt{\Omega_m (1 + z')^3 + \Omega_k (1 + z')^2 +
\Omega_{\Lambda}}}\right\}, \nonumber
\end{eqnarray}
where $H_0$ is the Hubble coefficient at $t_o$:
\begin{equation}\label{2.18}
H_0 = \left.S,_t/S\right|_{t = t_o}
\end{equation}
and the three dimensionless parameters
\begin{equation}\label{2.19}
\left(\Omega_m, \Omega_k, \Omega_{\Lambda}\right) \df \frac 1 {3{H_0}^2}
\left.\left(\frac {8\pi G \rho_0} {c^2}, - \frac {3 k} {{S_0}^2}, -
\Lambda\right)\right|_{t = t_o}
\end{equation}
obey $\Omega_m + \Omega_k + \Omega_{\Lambda} \equiv 1$ ($\rho_0$ is the current
mean mass density in the Universe and $S_0 = S(t_o)$). This formula applies also
with $\Omega_k < 0$ ($\sinh ({\rm i} x) \equiv {\rm i} \sin x$) and $\Omega_k
\to 0$. In the last case (\ref{2.17}) simplifies to
\begin{equation}\label{2.20}
D_L(z) = \frac {1 + z} {H_0} \int_0^z \frac {{\rm d} z'} {\sqrt{\Omega_m (1 +
z')^3 + \Omega_{\Lambda}}},
\end{equation}
where now $\Omega_m + \Omega_{\Lambda} \equiv 1$.

Note that the time coordinate $t$ used here is related to the physical time
$\tau$ (measured, for example, in years) by $t = c \tau$. Therefore, the Hubble
parameter $H_0$ defined in (\ref{2.18}) is related to the quantity ${\cal H}_0$
named ``Hubble constant'' in astronomical tables by
\begin{equation}\label{2.21}
H_0 = {\cal H}_0 / c.
\end{equation}

\section{Apparent horizons in the L--T and Friedmann models}\label{AHs}

\setcounter{equation}{0}

A general definition of an apparent horizon is given in Ref. \cite{HaEl1973}. In
application to the L--T models, one deals with a simpler situation
\cite{KrHe2004}, \cite{PlKr2006}. An apparent horizon (AH) is the boundary of a
region of spacetime, in which all bundles of null geodesics converge (have
negative expansion scalar -- for a model collapsing toward a final singularity)
or diverge (have positive expansion scalar -- for a model expanding out of a Big
Bang). The first kind of AH is called the future AH, the second one -- the past
AH. In what follows, only the past AHs will appear and the adjective ``past''
will be dropped.

The AH of the central observer is a locus where $R$, calculated along a
past-directed null geodesic given by (\ref{2.12}), changes from increasing to
decreasing, i.e., where
\begin{equation}\label{3.1}
\dr {} r R(t_{\rm ng}(r), r) = 0.
\end{equation}
This locus is given by \cite{PlKr2006}
\begin{equation}\label{3.2}
2M / R - 1 - \tfrac 1 3 \Lambda R^2 = 0.
\end{equation}
Equation (\ref{3.2}) has a solution for every value of $\Lambda$ (see Appendix
\ref{AHforeveryLambda}). Thus, as we proceed backward in time along the central
past light cone, the radius of the light cone first increases until the AH is
reached, then decreases, and this happens independently of the presence and sign
of $\Lambda$. The same applies to the Friedmann models \cite{Elli1971}.

{}From now on, $\Lambda = 0$ will be assumed for the L--T model, so the AH will
be at
\begin{equation}\label{3.3}
R = 2M = 2M_0 r^3.
\end{equation}
In the Friedmann limit this becomes
\begin{equation}\label{3.4}
S(t) = 2M_0 r^2.
\end{equation}

\section{The tilt of the matter velocity vector with respect to the
light cone in the $k = 0$ Friedmann model}\label{tilt}

\setcounter{equation}{0}

In Sections \ref{tilt} -- \ref{comments} the subcase $E = 0$ of the L--T model
will be considered, and the values of its parameters will be unrelated to
reality; they will be chosen so as to achieve the best visualisation. For a
radial null geodesic directed toward the center of symmetry, (\ref{2.12})
implies that the components of its tangent vector field obey
\begin{equation}\label{4.1}
k^t / k^r = - \left.R_{,r}\right|_{\rm ng}.
\end{equation}
This is a measure of the angle between the light cone and the flow lines of the
cosmic medium. In the Friedmann limit, with $r$ chosen as in (\ref{2.15}), the
above becomes
\begin{equation}\label{4.2}
\left.k^t / k^r\right|_{\rm F} = - \left.S(t)\right|_{\rm ng}.
\end{equation}
This determines the redshift via (\ref{2.16}).

\begin{figure}[h]
\hspace{-0.7cm}
\includegraphics[scale = 0.85]{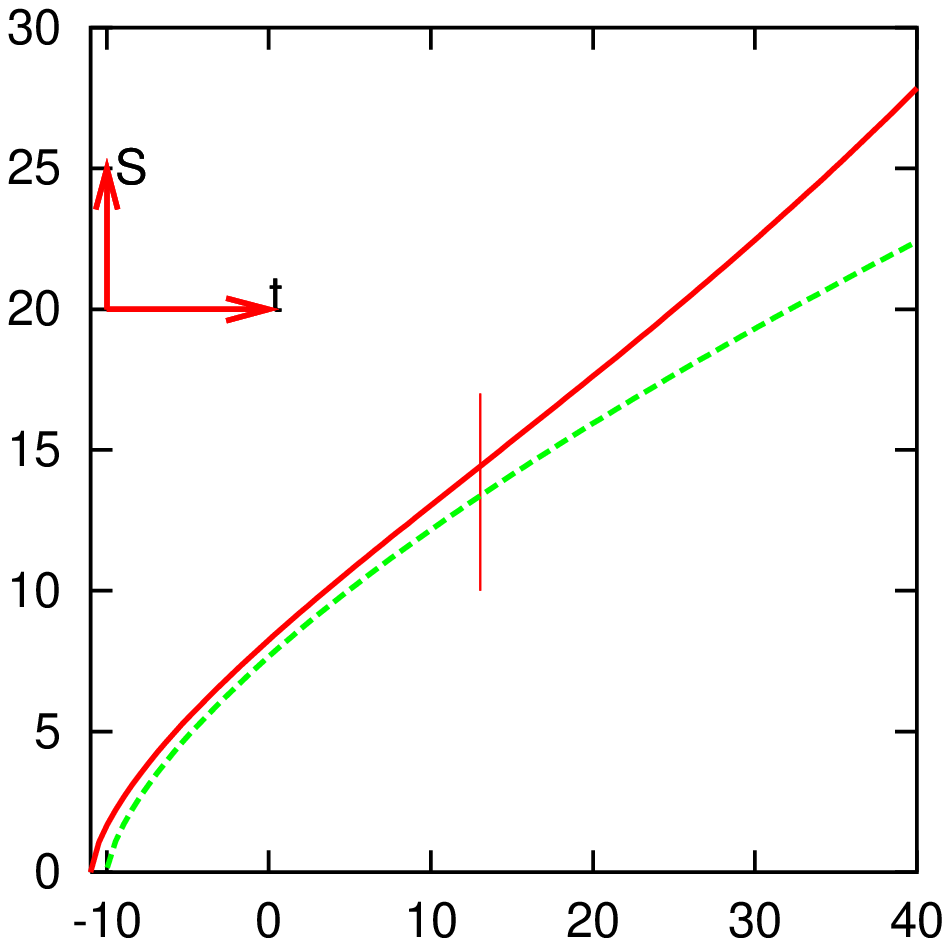}
 ${}$ \\[-4.5cm]
\hspace{17cm}
\includegraphics[scale = 0.5]{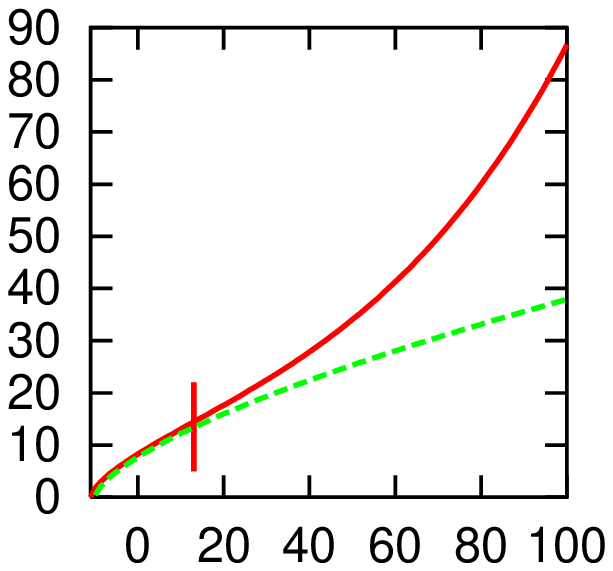}
\vspace{0.5cm} \caption{A comparison of the curves (\ref{4.4}) (the upper line)
and (\ref{4.5}) (the lower line). At the inflection point (marked by the
vertical bar) the accelerated expansion in (\ref{4.4}) sets in. The inset shows
the same curves over a longer period of time. The parameters are $(M_0, \Lambda,
t_{B0}, t_{B\Lambda}) = (1, -0.001, -10, -11)$. } \label{evolcompare}
\end{figure}

In the following, we will use the cosmologists' favourite Friedmann model, in
which $k = 0$ and $\Lambda < 0$. In this case, with $r$ defined as in
(\ref{2.9}), eq. (\ref{2.2}) becomes
\begin{equation}\label{4.3}
{S,_t}^2 = \frac {2M_0} S - \frac 1 3 \Lambda S^2.
\end{equation}
This has the elementary solution
\begin{equation}\label{4.4}
S_{\Lambda}(t) = \left(- \frac {6M_0} {\Lambda}\right)^{1/3} \sinh^{2/3} \left[\frac
{\sqrt {- 3 \Lambda}} 2 \left(t - t_{B\Lambda}\right)\right],
\end{equation}
where $t_{B\Lambda}$ is an arbitrary constant -- the time coordinate of the Big
Bang. For $\Lambda = 0$ the solution of (\ref{4.3}) is
\begin{equation}\label{4.5}
S(t) = \left(\frac {9M_0} 2\right)^{1/3} \left(t - t_{B0}\right)^{2/3},
\end{equation}
where $t_{B0}$ is another constant. Figure \ref{evolcompare} shows a comparison
of $S_{\Lambda}(t)$ and $S(t)$. (For the sake of easier comparison, the curve
(\ref{4.4}) in Fig. \ref{evolcompare} is shifted to $t_{B\Lambda} = -11$ instead
of $t_{B\Lambda} = -15$ used in most other figures.)

The following should be noted:

(1) The curve $S(t)$ is concave everywhere, while $S_{\Lambda}(t)$ is concave up
to the instant $t = t_i$, where
\begin{equation}\label{4.6}
t_i - t_{B\Lambda} = \frac 1 {\sqrt{- 3 \Lambda}} \ln \left(\frac {\sqrt{3} + 1}
{\sqrt{3} - 1}\right),
\end{equation}
and for $t > t_i$ becomes convex. At the inflection point $t = t_i$ the
accelerated expansion sets in.

(2) If $t_{B\Lambda} = t_{B0}$, then $S_{\Lambda}(t)$ and $S(t)$ are tangent at
$t = t_{B0}$.

(3) With $t_{B\Lambda} = t_{B0}$ we have, at any $t > t_{B0}$
\begin{equation}\label{4.7}
S_{\Lambda}(t) > S(t) \qquad {\rm and} \qquad S_{\Lambda,t} > S,_t.
\end{equation}

The basic measured quantity in cosmology is the Hubble parameter (\ref{2.18}).
Suppose, we want to compare the models (\ref{4.4}) and (\ref{4.5}), taking $H_0$
as given. Then, for $H_0$ being the same in both models, (\ref{2.18}) implies
\begin{equation}\label{4.8}
\frac {\sqrt{- 3 \Lambda}} 2\ \coth \left[\frac {\sqrt{- 3 \Lambda}} 2\ \left(t
- t_{B\Lambda}\right)\right] = \frac 1 {t - t_{B0}}.
\end{equation}
Since $\coth x > 1/x$ for all $x > 0$, Eq. (\ref{4.7}) implies
\begin{equation}\label{4.9}
t_{B\Lambda} < t_{B0},
\end{equation}
i.e. the Universe is older in the model (\ref{4.4}) than in (\ref{4.5}).

For later reference let us note that (\ref{3.2}) for the model (\ref{4.4}), in
the $(t, R)$ variables, has the form
\begin{equation}\label{4.10}
t = t_{B\Lambda} + \frac 2 {\sqrt{-3 \Lambda}}\ \ln\left(\frac {1 +
\sqrt{-\Lambda/3} R} {\sqrt{1 + \Lambda R^2/3}}\right).
\end{equation}

\section{The accelerated expansion}\label{accelerate}

\setcounter{equation}{0}

\begin{figure}
\begin{center}
\hspace{-1cm}
\includegraphics[scale = 0.65]{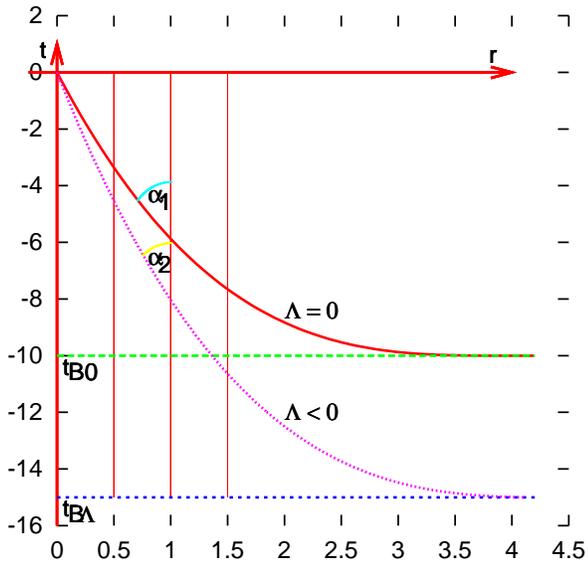}
\caption{The past null geodesic $t(r)$ for the metric (\ref{2.15}) with $k = 0 =
\Lambda$ (upper curve) and with $k = 0 > \Lambda$ (lower curve). The vertical
straight lines are world lines of the cosmic medium. We have $\alpha_2 <
\alpha_1$ everywhere. The observer is at $(t, r) = (0, 0)$; $t_{B\Lambda} =
-15$, other parameters are the same as in Fig. \ref{evolcompare}. This graph
does not faithfully show the radius of the intersection of the light cone with a
hypersurface of constant $t$; for that see Fig. \ref{conegeom}.}
\label{conecompare}
\end{center}
\end{figure}

With the $S(t)$ of (\ref{4.5}) the radial null geodesic equation for the metric
(\ref{2.15}) can be integrated:
\begin{equation}\label{5.1}
(t - t_{B0})^{1/3} = (t_o - t_{B0})^{1/3} - (M_0/6)^{1/3} r,
\end{equation}
where $(t, r)$ are the coordinates of the point on the geodesic and $t = t_o$ is
the instant of observation, at which $r = 0$. Figure \ref{conecompare} shows
this geodesic, compared with the null geodesic corresponding to (\ref{4.4}),
taking (\ref{4.9}) into account. With $\Lambda < 0$, the angle $\alpha_2$
between the geodesic and the flow lines of matter (which are the vertical
straight lines) is everywhere smaller than the corresponding angle $\alpha_1$
for $\Lambda = 0$, because of (\ref{4.7}) and (\ref{4.2}).

As we proceed back in time toward the Big Bang, more and more particles of the
cosmic matter are encompassed by the light cone. This is seen from (\ref{5.1}),
where $r(t)$ is decreasing in $t \in [t_{B0}, t_o]$. However, $r(t_{B0})$ is
finite,\footnote{The matter particle that leaves the Big Bang at $r(t_{B0})$ is
at the particle horizon \cite{Rind1956}, \cite{PlKr2006} at $t = t_o$.}
\begin{equation}\label{5.2}
r(t_{B0}) = \left[6(t_o - t_{B0})/M_0\right]^{1/3},
\end{equation}
i.e. the mass within the light cone is finite at the Big Bang (but $r(t_{B0})$
increases as $t_o$ increases.) The same is true for the $S_{\Lambda}$ of
(\ref{4.4}): because of (\ref{4.7}) we have
\begin{equation}\label{5.3}
r(t_{B\Lambda}) = \int_{t_{B\Lambda}}^{t_o} \frac {{\rm d} t} {S_{\Lambda}} <
\int_{t_{B\Lambda}}^{t_o} \frac {{\rm d} t} S < \infty.
\end{equation}

\begin{figure}
\begin{center}
\includegraphics[scale = 0.6]{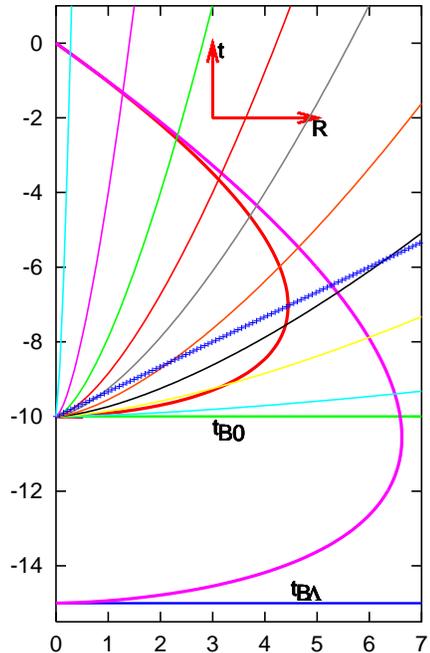}
\caption{The geodesic radius $R = rS(t(r))$ of the null cones from Fig.
\ref{conecompare} as a function of $t$. The curves fanning out of the point $(t,
R) = (-10, 0)$ are images of the vertical straight lines of Fig.
\ref{conecompare}. Each one of them has a different value of $r$. Note the
maximal value of $r$, beyond which the radius of the cone decreases toward the
Big Bang -- this is where the light cone intersects the past apparent horizon of
(\ref{4.5}), shown as the jagged straight line.} \label{conegeom}
\end{center}
\end{figure}

Figure \ref{conecompare} does not correctly display the spatial radius of the
light cones. It gives the illusion that the radius becomes ever larger toward
the Big Bang. This is not the case. With $k = 0$, the invariant radius of the
light cone at time $t$ is $R \df rS(t(r))$, where $t(r)$ is the function implied
by (\ref{5.1}) or its $\Lambda < 0$ counterpart. Figure \ref{conegeom} shows the
graphs of $R$ against $t$ along the light cones of the models (\ref{4.4}) and
(\ref{4.5}). As is seen, when we proceed toward the past, the radius of the
light cone increases at first, but acquires a maximum at a certain instant and
then decreases to zero as the Big Bang is approached.\footnote{The past light
cones of the L--T models with $\Lambda = 0$ have the same property. It is this
feature that was mistaken for a ``pathology'' and named ``critical point'' in
Ref. \cite{VFWa2006}.} The maximum is at the intersection of the light cone with
the past apparent horizon (AH) \cite{PlKr2006,KrHe2004}. The general equation of
AH, (\ref{3.4}), for the model (\ref{4.5}) reduces to $t = t_B + (2/3)R$, this
line is also shown in Fig. \ref{conegeom}.

Figure \ref{conegeom} also shows the flow lines of matter for the model
(\ref{4.5}) in the $(t, R)$ variables. They are all convex because the functions
$R(t)$ that they represent all have $R,_{tt} < 0$ (decelerated expansion).

\begin{figure}[h]
\begin{center}
\includegraphics[scale = 0.5]{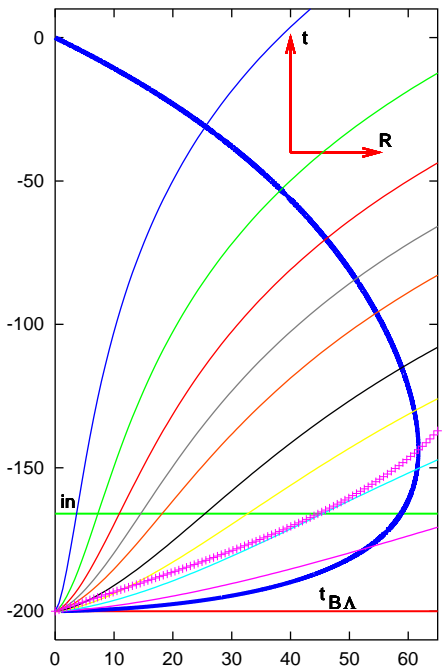}
\caption{The geodesic radius $R = rS(t(r))$ of the null cone corresponding to
(\ref{4.4}) as a function of $t$, and a collection of world lines of the cosmic
medium corresponding to different values of $r$. The horizontal line marked
``in'' is where all the world lines have their inflection points. The jagged
curve is the AH given by (\ref{4.10}). The values of the parameters are $(M_0,
\Lambda, t_{B\Lambda}) = (1, -0.0005, -200)$.}  \label{lamconegeometry}
\end{center}
\end{figure}

The inflection points of the flow lines for the model (\ref{4.4}), where the
accelerated expansion begins, with the parameter values used in Fig.
\ref{conegeom}, lie far to the future of the observer position $(t, R) = (0,
0)$. Therefore, for comparison, Fig. \ref{lamconegeometry} shows the
corresponding picture for the model (\ref{4.4}), with the parameter values
suitably adapted. It also shows the AH for this model, calculated from
(\ref{4.10}).

The observer does not know, which spacetime he/she is in, and only collects
light signals from the light cone. For the purpose of comparing the observations
carried out in the background of the model of (\ref{4.4}) with those carried out
in the background of (\ref{4.5}), we have to imagine the light cone of
(\ref{4.4}) being mapped into the light cone of (\ref{4.5}) in such a way that
the identity of the cosmic particles and the angle $\alpha_2$ (which is a
measure of the velocity of expansion) are preserved. To preserve the identity
means to move each point of the lower curve of Fig. \ref{conecompare} into the
upper curve along a vertical straight line.

\begin{figure}[h]
\vspace{-0.5cm}
\hspace{-0.5cm}
\includegraphics[scale = 0.85]{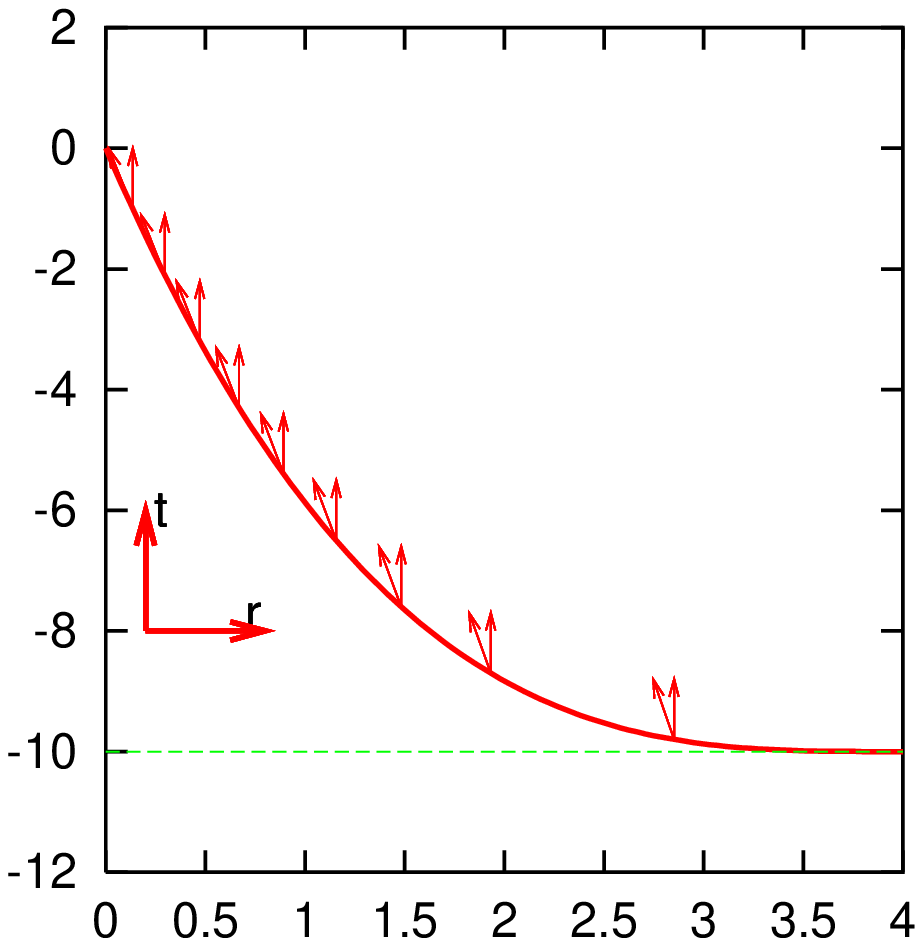}
${ }$ \\[-8.5cm]
\hspace{10.5cm}
\includegraphics[scale = 0.6]{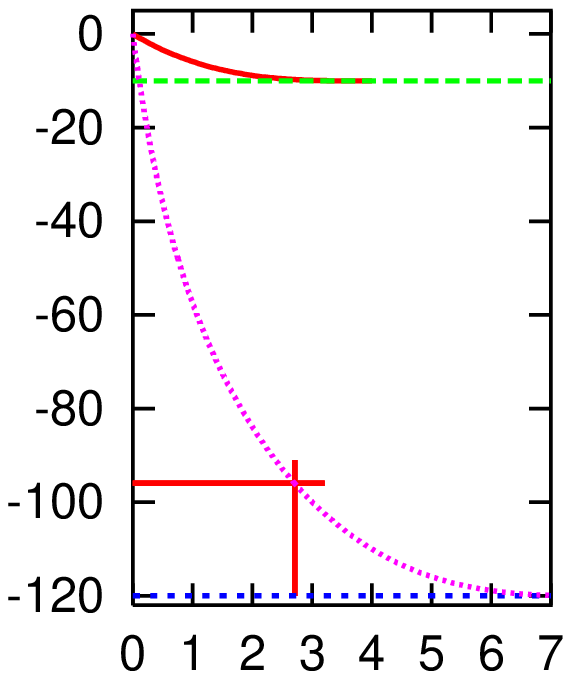}
\vspace{3cm}
\caption{When the observer at $(t, r) = (0, 0)$ interprets the redshift
observations against the background of the model (\ref{4.4}), the flow lines of
cosmic matter are tilted toward the light cone by more than was the case in
(\ref{4.5}). This excess tilt is a measure of the `accelerated expansion'. The
horizontal line at the bottom marks the time of the Big Bang for (\ref{4.5}).
The value of $t_{B\Lambda}$ is $-120$; other parameters are the same as in Fig.
\ref{evolcompare}. The inset shows Fig. \ref{conecompare} with $t_{B\Lambda}$
changed from $-15$ to $-120$. The crossing straight lines in the inset mark the
inflection point, where accelerated expansion begins.} \label{linecompare}
\end{figure}

Figure \ref{linecompare} shows the result of such a mapping. The $t_{B\Lambda}$
in it is $- 120$, so, by (\ref{4.6}), the accelerated expansion begins at $(t,
r) \approx (-96.0, 2.708)$. With (\ref{4.5}), all the flow lines have vertical
tangents, as in Fig. \ref{conecompare}. With (\ref{4.4}), the flow lines tilt
away from the vertical, more and more toward the light cone as $t$ increases.
The observer concludes that in the model given by (\ref{4.4}) the expansion rate
of the Universe increases with time.

\section{Explaining away the ``accelerated expansion'' by a nonsimultaneous Big
Bang}\label{alternative}

\setcounter{equation}{0}

It is shown below that the function $(k^t/k^r)(r)$ along the past light cone of
the observer implied by (\ref{4.4}) can be obtained using the $E = 0$ L--T
model. In order to calculate it, the corresponding null geodesic equation for
(\ref{2.15}) with (\ref{4.4}) is first solved:
\begin{equation}\label{6.1}
\dr t r = - \left(\frac {6M_0} {- \Lambda}\right)^{1/3} \sinh^{2/3}\left[\frac
{\sqrt{- 3 \Lambda}} 2 \left(t - t_{B\Lambda}\right)\right].
\end{equation}
The solution (found numerically and shown in Fig. \ref{conecompare}) will be
denoted $t = t_F(r)$. When it is substituted in (\ref{4.4}), it determines
$(k^t/k^r)_F(r)$ via (\ref{4.2}).

The corresponding $k^t/k^r$ in the L--T model (\ref{2.6}) is found from
(\ref{2.12}), which, with $r$ chosen as in (\ref{2.9}), reads
\begin{eqnarray}\label{6.2}
\dr t r &=& \left(\frac {9M_0} 2\right)^{1/3} \left\{- \left[t -
t_B(r)\right]^{2/3}\right. \nonumber \\
&+& \left.\frac 2 3\ r \left[t - t_B(r)\right]^{-1/3} t_{B,r}\right\}.
\end{eqnarray}
The solution of (\ref{6.2}) will be denoted $t = t_{LT}(r)$.

The same function $(k^t/k^r)(r)$ along the past light cone in both models will
thus follow when
\begin{equation}\label{6.3}
\left(\dr t r\right)_F = \left(\dr t r\right)_{LT}.
\end{equation}
This means that $t_F$ and $t_{LT}$ will coincide at the observer's position when
$t_F = t_{LT}$ everywhere on the cone. Consequently, $t_F(r)$ must be found from
(\ref{6.1}), then substituted for $t$ in (\ref{6.2}). The result can be written
as
\begin{eqnarray}\label{6.4}
\dr {t_B} r &=& \frac 3 {2r} \left\{\left(\frac 2 {9M_0}\right)^{1/3}
\left[t_F(r) -
t_B(r)\right]^{1/3} \dr {t_F} r\right. \nonumber \\
&&\ \ \ \ \ + \left.t_F(r) - t_B(r)\right\},
\end{eqnarray}
\noindent where $\dril {t_F} r$ is given by (\ref{6.1}). A necessary condition
for $t_{B,r}$ to be finite at $r = 0$ is that the expression in braces tends to
zero when $r \to 0$. This will happen if
\begin{equation}\label{6.5}
\lim_{r \to 0} \frac {\sinh \left\{\frac {\sqrt{-3 \Lambda}} 2 \left[t_F(r) -
t_{B\Lambda}\right]\right\}} {\frac {\sqrt{-3 \Lambda}} 2 \left[t_F(r) -
t_B(r)\right]} = 1.
\end{equation}
This determines the value of $t_B(0)$:
\begin{eqnarray}\label{6.6}
t_B(0) &=& t_F(0) - \frac 2 {\sqrt{-3 \Lambda}} \sinh \left\{\frac {\sqrt{-3
\Lambda}} 2 \left[t_F(0) - t_{B\Lambda}\right]\right\} \nonumber \\
&<& t_{B\Lambda}.
\end{eqnarray}
Note that $[t_F(0) - t_B(0)]$ increases when $|\Lambda|$ increases. With
(\ref{6.6}) fulfilled, (\ref{6.4}) implies
\begin{eqnarray}\label{6.7}
&& \lim_{r \to 0} \dr {t_B} r = \frac 1 2 \left(\frac {6 M_0} {-
\Lambda}\right)^{1/3} \nonumber \\
&\times& \left\{\cosh \left[\frac {\sqrt{-3 \Lambda}} 2 \left(t_F(0) -
t_{B\Lambda}\right)\right] - 1\right\} \nonumber \\
&\times& \sinh^{2/3}\left[\frac {\sqrt{-3 \Lambda}} 2 \left(t_F(0) -
t_{B\Lambda}\right)\right] > 0.
\end{eqnarray}
This implies $\lim_{r \to 0}\rho,_r > 0$ for the $\rho$ of (\ref{2.8}), which
relates in two ways to problems considered in the literature:

1. The property $\rho,_r \neq 0$ at the center was called ``weak singularity''
\cite{VFWa2006}. However, this is not a singularity in the sense of any
definition used in relativity \cite{KHBC2010}.

2. When $\rho,_r > 0$ at the center, the density increases with distance from the
center, i.e. there is a void around the center. Several astrophysicists believe
that the presence of this void is a necessary feature of any L--T model used to
mimic accelerated expansion (see references in Sec. \ref{LTcosmology}). The model
considered in our Secs. \ref{duplicate} -- \ref{numcone} is a counterexample to
this belief.

\begin{figure}[h]
\begin{center}
\vspace{-0.1cm}
\includegraphics[scale=0.6]{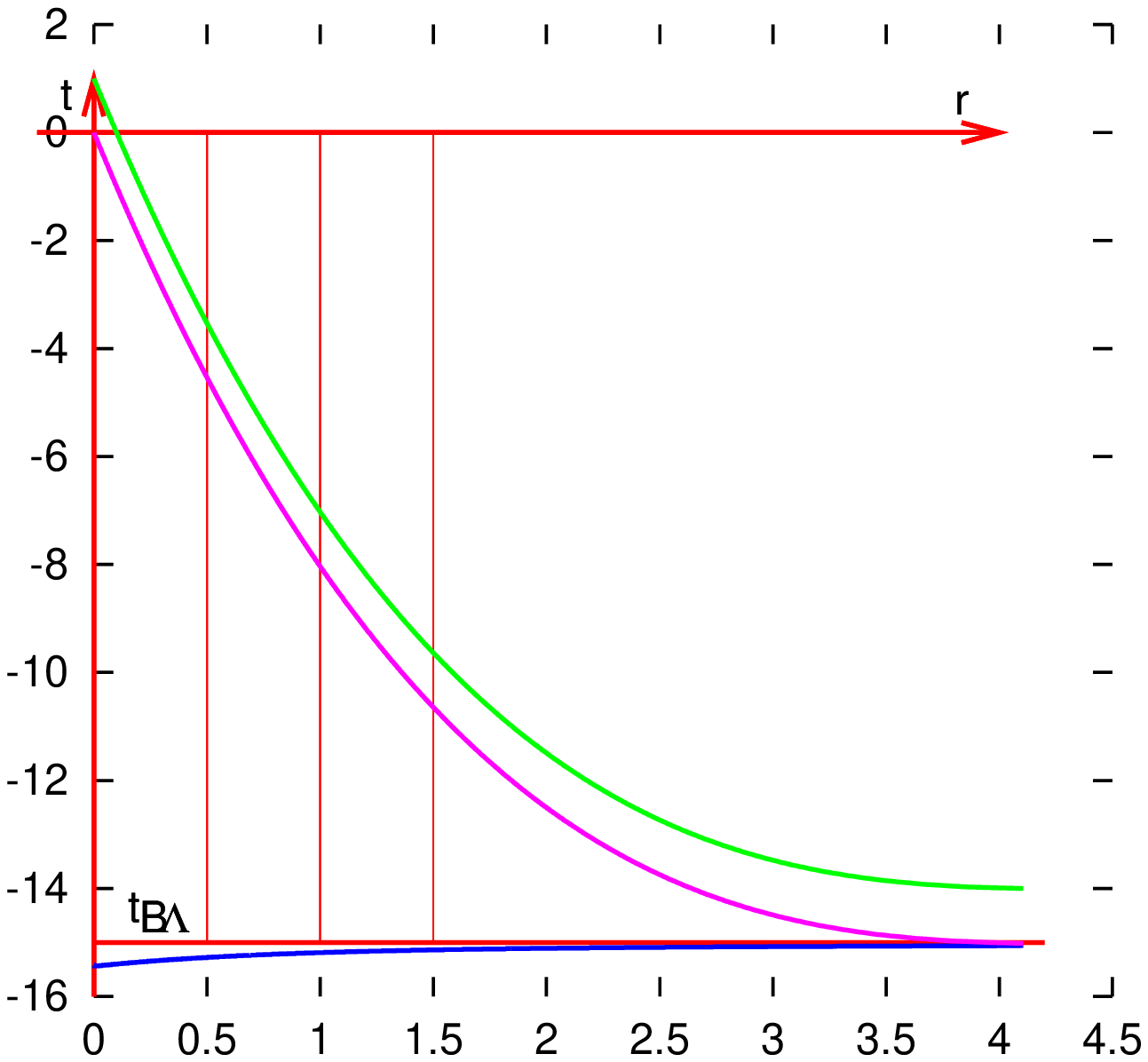}
 ${}$ \\[-6.3cm]
\hspace{2.5cm}
\includegraphics[scale = 0.5]{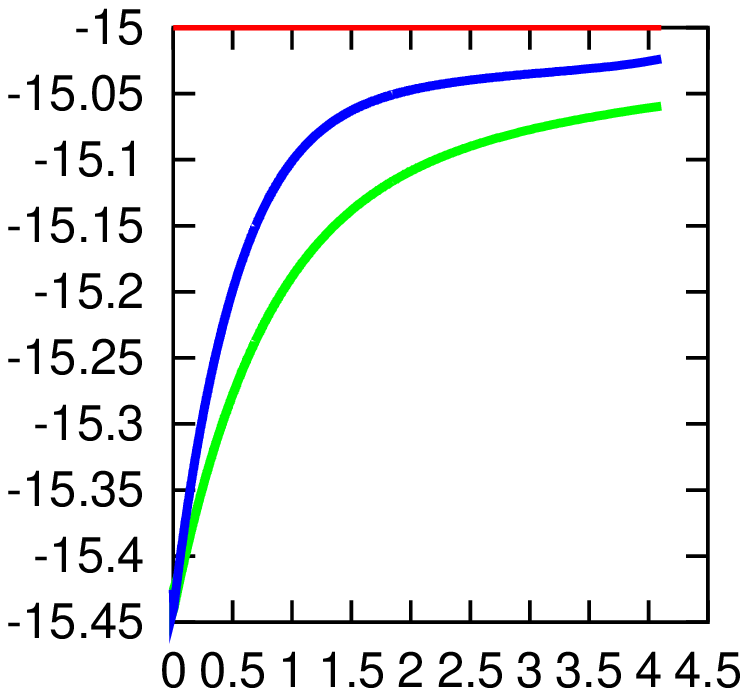}
\vspace{2cm} \caption{{\bf Lowest curve:} The function $t_B(r)$ defined by
(\ref{6.4}) and (\ref{6.1}). {\bf Middle curve:} The L--T light cone calculated
from (\ref{6.2}) with $t_B(r)$ as in the lowest curve. {\bf Upper curve:} The
$\Lambda < 0$ light cone from Fig. \ref{conecompare}, shifted by $\Delta t = 1$
upwards, included for comparison. The inset shows a closeup view of the time
interval between the shell crossing (upper curve) and the Big Bang (lower
curve).} \label{drawbang}
\end{center}
\end{figure}

Figure \ref{drawbang} shows the graph of $t_B(r)$ calculated from (\ref{6.4}),
the corresponding past light cone for the central observer, and the $\Lambda <
0$ past light cone from Fig. \ref{conecompare}, included for comparison. The two
light cones coincide up to numerical errors $\Delta t \approx 0.015$. Assuming
that $- t_{B\Lambda} = 15$ represents the age of the Universe $T = 13.819 \times
10^9$ y \cite{Plan2013}, this error translates to $\Delta t = 10^{-3} T = 1.38
\times 10^7$ y.

Since in all the models comoving coordinates were used, the flow lines of matter
are vertical straight lines in every case. Therefore, identical light cones for
the two models mean identical functions $k^t/k^r$ for both.

The inset in Fig. \ref{drawbang} shows the shell crossing (given by
(\ref{2.11})), which is in this case inevitable, since $t_B(r)$ is increasing
all the way. (At the scale of the main figure, the shell crossing would coincide
with the Big Bang.)

Since the example discussed up to now was not meant to reflect any real
measurements done in astronomy, the question whether the shell crossings pose a
serious problem is irrelevant. But, for the sake of completeness, let us note
the following. The biggest time-difference between the shell crossing and the
Big Bang is 0.0885 time units used in the figure, while the time-difference at
the right margin is 0.0354. This translates to $8.15 \times 10^7$ y and $3.26
\times 10^7$ y, respectively. This is to be compared with $t - t_B \approx 3.5
\times 10^5$ years for the recombination epoch -- so, clearly, this is not a
realistic model of our Universe.

\begin{figure}[h]
\begin{center}
\includegraphics[scale = 0.6]{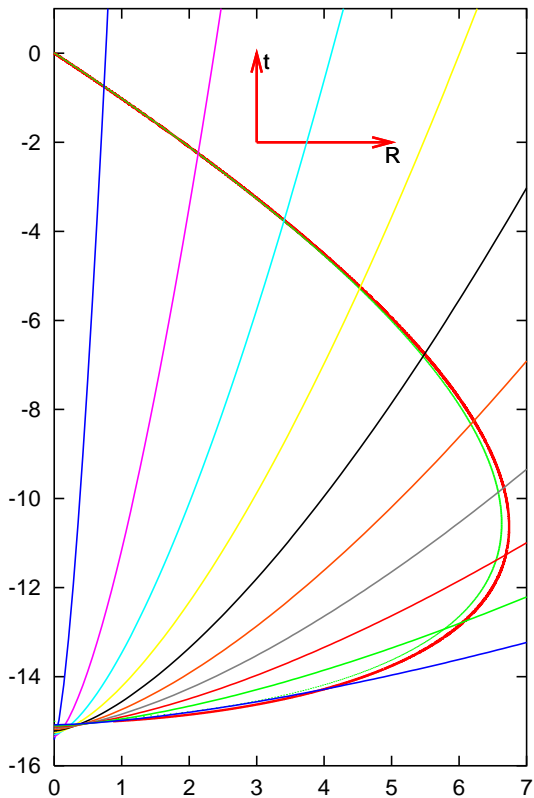}
\caption{The geodesic radius $R(t(r))$ of the null cone defined by (\ref{6.2})
and (\ref{6.4}) as a function of $t$, and a collection of flow lines of the
cosmic medium corresponding to different values of $r$. The Big Bang is now a
finite segment of the $t$-axis. Note the intersections of the flow lines in the
vicinity of their origins -- they are images of shell crossings. A closeup view
of the shell crossings is shown in Fig. \ref{ltworldlinesfocus}. The second
curve seen in the neighbourhood of maximal $R$ is a copy of the light cone from
Fig. \ref{lamconegeometry}. The two cones do not coincide in the $(t, R)$
variables.} \label{ltworldlines}
\end{center}
\end{figure}

\begin{figure}[h]
\begin{center}
\includegraphics[scale = 0.7]{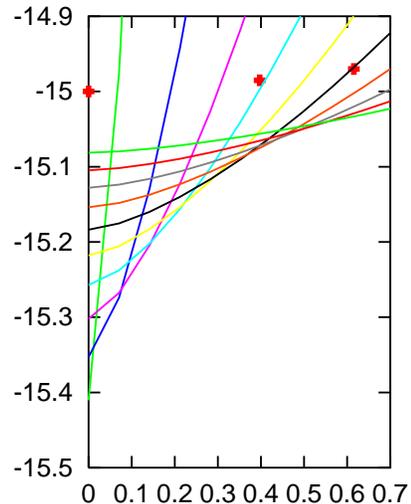}
\caption{A closeup view of the region of shell crossings in Fig.
\ref{ltworldlines}. The crosses mark the $\Lambda < 0$ Friedmann null cone.}
\label{ltworldlinesfocus}
\end{center}
\end{figure}

It is interesting to transform Fig. \ref{drawbang} to the variables $(t, R)$, in
analogy to Fig. \ref{lamconegeometry}. The result of the transformation is shown
in Fig. \ref{ltworldlines}. Now the Big Bang is no longer a single point, but a
segment of the $t$-axis. This reflects the fact that the Big Bang occurs at
different times for different flow lines. The flow lines no longer have a common
origin, and they intersect in the vicinity of their origins. The intersections
are images of the shell crossings, shown in closeup view in Fig.
\ref{ltworldlinesfocus}.

The light cone in Fig. \ref{ltworldlines} does not extend to the $R = 0$ line
because of numerical errors. They cause that the light cone in Fig.
\ref{drawbang} ends at $r \approx 4$, where it has not yet met the Big Bang set,
so $R$ is not yet zero there, and the gap is magnified in the transformation.
For the same reason, the two light cones from Fig. \ref{drawbang} coincide with
a smaller precision after the transformation -- the image of the $\Lambda < 0$
Friedmann cone is seen in the vicinity of the maximal radius in Fig.
\ref{ltworldlines}.

\section{Comments}\label{comments}

\setcounter{equation}{0}

Since solving (\ref{6.1}) only requires calculating an integral of a function of
$r$ which is evidently integrable, the solution exists for every $t_{B\Lambda}$.
The same is true for (\ref{6.2}): the $t_B(r)$ determined by it exists for every
$t_F(r)$. However, the solution of (\ref{6.1}) defines {\em a single light cone}
of (\ref{2.15}). {\em The same} L--T model will not mimic all light rays in
(\ref{4.4}) reaching a given observer. The time evolution of the L--T model with
$\Lambda = 0$ is different from that of the $\Lambda$CDM model, and the two can
be distinguished by observations that are sensitive to the dynamics of the
Universe, and not just to a momentary ``snapshot''. Examples of effects that
depend on the time-evolution are redshift drift \cite{QABC2012} and
non-repeatability of light paths \cite{KrBo2011,Kras2011,Kras2012}.

The function $t_B(r)$ in Fig. \ref{drawbang} is increasing, and $t_B(r) <
t_{B\Lambda}$ at all $r$. To get an understanding why this is so, let us observe
the following. The L--T model of (\ref{2.6}) expands by the same law as the $k =
0$ Friedmann model with $\Lambda = 0$. Because of (\ref{4.7}) the function
$(k^t/k^r)_F = -S_{\Lambda}(t)$ decreases faster with $t$ than $(k^t/k^r)_{LT} =
- R,_r$ at $r = 0$. Hence, in order to {\em slow down} to the same rate of
decrease as $(- S_{\Lambda})$, the function $(- R,_r)$ needs more time, so
$t_B(0)$ must precede $t_{B\Lambda}$. With the age of the Universe $(t(r) -
t_{B\Lambda})$ decreasing along the past light cone, $t_B - t_{B\Lambda}$ must
also decrease, so $t_B(r)$ must be increasing. In order to obtain $t_B(r) >
t_{B\Lambda}$, one needs to consider a quantity that either decreases slower or
increases faster in the $\Lambda$CDM model than in L--T.

Since just one of the two arbitrary functions in the L--T model suffices to
mimic accelerated expansion, it is natural to suppose that with both functions,
$E(r)$ and $t_B(r)$, being arbitrary, the L--T model can be adapted to two sets
of observations. In Ref. \cite{CBKr2010} it was explicitly demonstrated that
this is indeed possible for the pairs (angular diameter distance -- mass density
in the redshift space) and (angular diameter distance -- expansion rate).

Note how eq. (\ref{6.2}) displays an instability of the Friedmann model with
respect to the L--T perturbation.\footnote{This was first observed by Szekeres
\cite{Szek1980}, and discussed in more detail by Hellaby and Lake
\cite{HeLa1984}, see also Ref. \cite{PlKr2006}.} In the Friedmann limit, we have
$t_{B,r} = 0$, so $\lim_{t \to t_B} \dril t r = 0$, i.e., in the comoving
coordinates, the tangent to each null geodesic becomes horizontal at the Big
Bang. However, in the L--T model, at every $r > 0$ where $t_{B,r} \neq 0$, we
have $\lim_{t \to t_B} \left|\dril t r\right| = \infty$, i.e. the tangent to the
null geodesic is vertical. The only exceptions are points in which $t_{B,r} =
0$, where the said tangent is horizontal even in L--T. Thus, since in our L--T
model the current observer's light cone is the same as in a Friedmann model,
this light cone must be horizontal at $t = t_B$. This means that the observer
who carried out this construction must live in a special epoch: that, in which
her past light cone intersects with the extremum/inflection of the Big Bang set.
This should not be disturbing from the point of view of astrophysics, for the
following reasons:

1. The dust models do not apply just after the Big Bang -- the pressure cannot
be assumed zero at those early times. They begin to apply no earlier than after
last scattering. Considering the light cones up to the Big Bang was a geometric
exercise, whose results are not to be taken as implications for our physical
Universe.

2. Light from objects that might have existed before last scattering is not
observed. Hence, we have no observational clues as to the state of the Universe
prior to that epoch. (The situation might improve when neutrinos and
gravitational waves from the early Universe can be registered, but this will
happen in the future, perhaps distant future.) Also, there is a long gap between
the highest-redshift objects observed so far ($z \approx 10$)\footnote{For a
somewhat outdated summary on the objects with highest redshifts see Ref.
\cite{McMa2005}.} and the last scattering epoch ($z \approx 1089$
\cite{Luci2004}); we have no direct information from that segment of our past
light cone. Consequently, the attempt to reconstruct our whole past light cone
up to its contact with the Big Bang is excessively ambitious -- the result is
not observationally testable.

3. For simplicity, the adequacy of the $\Lambda$CDM model was not discussed
here, and the values of its parameters were taken for granted. However, in order
to test the L--T model against observations in earnest, one would have to use it
in the analysis of observational data from the beginning to the end. The L--T
model should be adapted directly to the observational data, and not to the
parameters of the best-fit Friedmann model. Such an analysis still remains to be
done.

The peculiar properties of the L--T light cone will be present also in the model
discussed further on, see eqs. (\ref{11.2}) and (\ref{11.4}).

The discussion up to this place was presented for illustrative purposes. It is
related to astrophysics indirectly, but is free from numerical complications.
{}From the next section on, a more realistic example will be described.

\section{Duplicating the luminosity distance -- redshift relation using the L--T
model with $\Lambda = 0$}\label{duplicate}

\setcounter{equation}{0}

Now it will be shown how the luminosity distance -- redshift relation of the
$\Lambda$CDM model (our eq. (\ref{2.20})) can be duplicated using the L--T model
with $\Lambda = 0$. The reasoning below was inspired by Iguchi et al.
\cite{INNa2002}.

To duplicate (\ref{2.20}) using the $\Lambda = 0$ L--T model means, in view of
(\ref{2.14}) and (\ref{2.20}), to require that
\begin{equation}\label{8.1}
R(t_{\rm ng}(r), r) = \frac 1 {H_0 (1 + z)} \int_0^z \frac {{\rm d} z'}
{\sqrt{\Omega_m (1 + z')^3 + \Omega_{\Lambda}}}
\end{equation}
holds along the past light cone of the central observer, where $H_0, \Omega_m$
and $\Omega_{\Lambda}$ have the values determined by current observations,
$t_{\rm ng}(r)$ is the function determined by (\ref{2.12}) and $z(r)$ is
determined by (\ref{2.13}). Let
\begin{equation}\label{8.2}
{\cal D}(z) \df \int_0^z \frac {{\rm d} z'} {\sqrt{\Omega_m (1 + z')^3 +
\Omega_{\Lambda}}}.
\end{equation}
Note that ${\cal D}(0) = 0$, ${\cal D}(z) > 0$ at all $z > 0$ and ${\cal D},_z >
0$ at all $z \geq 0$, but $\lim_{z \to \infty}{\cal D}(z)$ is finite, since, for
$\Omega_{\Lambda} > 0$ (as is the case in the $\Lambda$CDM model) at all $z$ we
have
\begin{eqnarray}\label{8.3}
&& {\cal D}(z) < \int_0^z \frac {{\rm d} z'} {\sqrt{\Omega_m (1 + z')^3}}
\nonumber \\
&& = \frac 2 {\sqrt{\Omega_m}} \left(1  - \frac 1 {\sqrt{1 + z}}\right) < \frac
2 {\sqrt{\Omega_m}} < \infty.
\end{eqnarray}

Unlike in the RW models, light emitted at the Big Bang of an L--T model and
reaching an observer is in general infinitely blueshifted, i.e. $z_{\rm BB} =
-1$, except when $t_{B,r} = 0$ at the emission point \cite{Szek1980},
\cite{HeLa1984}, \cite{PlKr2006} -- then it may have infinite redshift. As
follows from (\ref{8.1}) and (\ref{8.3}), at the Big Bang, where $R = 0$, $z \to
\infty$ must hold. This implies that, just like in the previous example in Sec.
\ref{comments}, $t_{B,r} \to 0$ at the emission point of the ray (\ref{8.1})
should hold. Note also, from (\ref{8.1}) and (\ref{8.3}) again, that at the Big
Bang, where $z \to \infty$, the following is true
\begin{equation}\label{8.4}
\lim_{z \to \infty} \left\{R_{\rm ng} (1 + z)\right\} = C_0 < \infty.
\end{equation}

\section{Locating the apparent horizon}\label{locateAH}

\setcounter{equation}{0}

Recall: at the AH $(\dril {} r) \left.R\right|_{\rm ng} = 0$ \cite{KrHe2004},
\cite{PlKr2006}. Thus, differentiating (\ref{8.1}) by $r$, one obtains
\begin{equation}\label{9.1}
\left(A_1 \dr z r\right)_{\rm AH} = 0,
\end{equation}
where
\begin{equation}\label{9.2}
A_1 \df {\cal D} - \frac {1 + z} {\sqrt{\Omega_m (1 + z)^3 + \Omega_{\Lambda}}}.
\end{equation}
Suppose, for a moment, that $\left.\dril z r\right|_{\rm AH} = 0$. In
consequence of (\ref{2.13}), this would mean $\left.R,_{tr}\right|_{\rm AH} =
0$. It is shown in Appendix \ref{whereAH} that this equation forces a relation
between $M$, $E$ and $t_B$, thus reducing the number of arbitrary functions to
2. So, $\left.R,_{tr}\right|_{\rm AH}$ is zero only in those special
cases,\footnote{The remark in Ref. \cite{INNa2002}, made after their (3.1),
which implies that the locus of $R,_{tr} = 0$ coincides with $R = 2M$, is thus
incorrect.} while the general conclusion from (\ref{9.1}) is
\begin{equation}\label{9.3}
\left.A_1\right|_{\rm AH} = 0.
\end{equation}
Note that this equation does not refer to the L--T model.

\begin{figure}[h]
\begin{center}
\includegraphics[scale = 0.5]{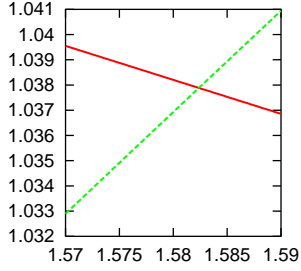}
\caption{Graphical solution of the equation $A_1 = 0$ with $A_1$ given by
(\ref{9.2}) and $(\Omega_m, \Omega_{\Lambda}) = (0.32, 0.68)$, as in Ref.
\cite{Plan2013}. The increasing function is ${\cal D}(z)$, the decreasing
function is $(1 + z) / {\sqrt{\Omega_m (1 + z)^3 + \Omega_{\Lambda}}}$. The
intersection of the curves is at the apparent horizon, where $z \approx 1.583$,
${\cal D} \approx 1.038$.} \label{drawdfun}
\end{center}
\end{figure}

With $\Omega_m$ and $\Omega_{\Lambda}$ given, the equation $A_1 = 0$ can be
solved for $z$. Using the values $(\Omega_m, \Omega_{\Lambda}) = (0.32, 0.68)$
as in Ref. \cite{Plan2013}, Fig. \ref{drawdfun} shows that $z \approx 1.583$ on
the AH.

The equation of the AH, (\ref{3.3}), may be written, using (\ref{8.1}),
(\ref{8.2}) and (\ref{2.9}), also as
\begin{equation}\label{9.4}
r_{\rm AH} = \left[\frac {\cal D} {2M_0 H_0 (1 + z)}\right]^{1/3}_{\rm AH}.
\end{equation}
In what follows, it will be useful to define one more quantity that vanishes on
the AH, in consequence of (\ref{9.4}):
\begin{equation}\label{9.5}
A_2 \df \sqrt{\frac {2M_0 H_0 r^3 (1 + z)} {\cal D}} - 1.
\end{equation}

The equation of the AH can be written in yet another form. For the case $E > 0$,
from (\ref{2.5}) and (\ref{3.3}), we have
\begin{eqnarray}\label{9.6}
&& \left(t - t_B\right)_{\rm AH} = \\
&& \left\{\frac M {(2E)^{3/2}}\ \left[\sqrt{{\cal Y}^2 - 1} - \ln \left({\cal Y}
+ \sqrt{{\cal Y}^2 - 1}\right)\right]\right\}_{\rm AH}, \nonumber
\end{eqnarray}
where
\begin{equation}\label{9.7}
{\cal Y} \df 1 + 4E.
\end{equation}
For $2E/r^2 = -k =$ constant (\ref{9.6}) becomes
\begin{eqnarray}\label{9.8}
&& \left(t - t_B\right)_{\rm AH} = \frac {M_0} {(-k)^{3/2}}
\left\{\sqrt{\left(1 - 2kr^2\right)^2 - 1}\right. \nonumber \\
&& - \left.\ln \left[1 - 2kr^2 + \sqrt{\left(1 - 2kr^2\right)^2 -
1}\right]\right\}_{\rm AH}.\ \ \ \
\end{eqnarray}

\section{The numerical units}\label{numerunits}

\setcounter{equation}{0}

The following values are assumed here
\begin{equation}\label{10.1}
(\Omega_m, \Omega_{\Lambda}, H_0, M_0) = (0.32, 0.68, 6.71, 1)
\end{equation}
the first two after Ref. \cite{Plan2013}. The $H_0$ is $1/10$ of the
observationally determined value of the Hubble constant \cite{Plan2013}
\begin{equation}\label{10.2}
{\cal H}_0 = c H_0 = 67.1\ {\rm km/(s} \times {\rm Mpc}).
\end{equation}
It follows that $H_0$ is measured in 1/Mpc. Consequently, choosing a value for
$H_0$ amounts to defining a numerical length unit (call it NLU), which, with
(\ref{10.1}), obeys
\begin{equation}\label{10.3}
H_0 = \frac {67.1} {3 \times 10^5}\ \frac {\rm (km/s)/Mpc} {\rm (km/s)} = 6.71
\frac 1 {\rm NLU}.
\end{equation}
{}From here
\begin{equation}\label{10.4}
1\ {\rm NLU} = 3 \times 10^4\ {\rm Mpc}.
\end{equation}

With $\Omega_{\Lambda}$ and $H_0$ given by (\ref{10.1}) we obtain from
(\ref{2.19})
\begin{equation}\label{10.5}
- \Lambda = 3 \Omega_{\Lambda} {H_0}^2 = 91.849164 ({\rm NLU})^{-2}.
\end{equation}
Since our time coordinate is $t = c \tau$, where $\tau$ is measured in time
units, $t$ is measured in length units. So it is natural to take the NLU defined
in (\ref{10.4}) also as the numerical time unit (NTU). We take the following
approximate values for the conversion factors \cite{unitconver}:
\begin{eqnarray}\label{10.6}
1\ {\rm pc} &=& 3.086 \times 10^{13}\ {\rm km}, \nonumber \\
1\ {\rm y} &=& 3.156 \times 10^7\ {\rm s}.
\end{eqnarray}
The following relations result, using (\ref{10.4}):
\begin{eqnarray}\label{10.7}
&& 1\ {\rm NTU} = 1\ {\rm NLU} = 3 \times 10^4\ {\rm Mpc} \nonumber \\
&&= 9.26 \times 10^{23}\ {\rm km} = 9.8 \times 10^{10}\ {\rm y}.
\end{eqnarray}
Using this in (\ref{10.5}) we get
\begin{equation}\label{10.8}
- \Lambda = 1.02 \times 10^{-7}\ ({\rm Mpc})^{-2}.
\end{equation}
Finally, for the age of the Universe \cite{Plan2013}
\begin{equation}\label{10.9}
T = 13.819 \times 10^9\ {\rm y}
\end{equation}
we obtain
\begin{equation}\label{10.10}
T = \frac {13.819 \times 10^9} {9.8 \times 10^{10}}\ {\rm NTU} = 0.141\ {\rm
NTU}.
\end{equation}
The values (\ref{10.5}) and (\ref{10.10}) will be used for the model
(\ref{4.4}), with $t_{B\Lambda} = - T$.

As already mentioned below (\ref{2.9}), $M_0$ represents mass, but has the
dimension of length ($M_0 = G m_0/c^2$, where $m_0$ is measured in mass units).
The choice $M_0 = 1$ NLU made in (\ref{10.1}) simplifies all computations. The
associated mass unit $M_0 c^2/G \approx 10^{57}$ kg will not appear in any other
way than via $M_0$.

\section{The L--T model with $2E = - k r^2$ that duplicates the $D_L(z)$ of
(\ref{2.20})}\label{LTwithnonzeroE}

\setcounter{equation}{0}

The functional shape of $E$ might be determined by tying it to an additional
observable quantity, as was done in Ref. \cite{CBKr2010}. However, then the
equations defining $t_B$ and $E$ become coupled, and numerical handling becomes
instantly necessary. To keep things transparent, we will rather follow the
approach of Ref. \cite{INNa2002}, and take
\begin{equation}\label{11.1}
2E = - kr^2,
\end{equation}
where $k < 0$ is an arbitrary constant. This $E$ is the same as in the $k < 0$
Friedmann model. The $M$ will be chosen as in (\ref{2.9}). From (\ref{2.10}) we
have on the light cone
\begin{equation}\label{11.2}
\dr t r = - \frac {R,_r} {\sqrt{1 - k r^2}},
\end{equation}
where the general formula for $R,_r$ is (\cite{PlKr2006}, eq. (18.104))
\begin{eqnarray}\label{11.3}
R,_r &=& \left(\frac {M,_r} M - \frac {E,_r} E\right)R \nonumber \\
&+& \left[\left(\frac 3 2 \frac {E,_r} E - \frac {M,_r} M\right) \left(t -
t_B\right) - t_{B,r}\right] R,_t.\ \ \ \
\end{eqnarray}
Using (\ref{11.1}), (\ref{2.2}) and (\ref{2.9}) this simplifies to
\begin{equation}\label{11.4}
R,_r = \frac R r - r t_{B,r} \sqrt{\frac {2M_0 r} R - k}.
\end{equation}
Equation (\ref{11.4}) substituted in (\ref{11.2}) leads to the same conclusions
about the L--T light cone that were formulated in paragraph 4 of Sec.
\ref{comments}.

With (\ref{11.1}), eqs. (\ref{2.7}) become
\begin{eqnarray}
\cosh \eta &=& 1 - \frac {k R} {M_0 r}, \label{11.5} \\
t - t_B &=& \frac {M_0} {(- k)^{3/2}} (\sinh \eta - \eta). \label{11.6}
\end{eqnarray}
Equations (\ref{11.5}) -- (\ref{11.6}) will now be taken along a null geodesic,
i.e. the $t$ in (\ref{11.6}) will be the $t(r)$ defined by (\ref{11.2}), while
the $R$ in (\ref{11.5}) will be the $R_{\rm ng}$ from (\ref{8.1}). We thus have
from (\ref{11.6})
\begin{equation}\label{11.7}
\dr t r - \dr {t_B} r = \frac R {\sqrt{- k} r} \left.\dr \eta r\right|_{\rm ng},
\end{equation}
where, from (\ref{11.5})
\begin{equation}\label{11.8}
\left.\dr \eta r\right|_{\rm ng} = - \frac {kr} {\sqrt{k^2 {R_{\rm ng}}^2 -
2kM_0 r R_{\rm ng}}} \left(\frac {R_{\rm ng}} r\right),_r.
\end{equation}
Substituting for $\dril t r$ from (\ref{11.2}) and (\ref{11.4}), and for $R_{\rm
ng}$ from (\ref{8.1}), then using (\ref{9.2}) and (\ref{9.5}) as the definitions
of $A_1$ and $A_2$, we obtain from (\ref{11.7})
\begin{eqnarray}\label{11.9}
&& B_2 \left[\frac {\cal D} {H_0 r (1 + z)} - \sqrt{\left(A_2 + 1\right)^2 -
kr^2} \dr {t_B} r\right] \nonumber \\
&& = \frac {A_1 \sqrt{1 - kr^2}} {H_0 (1 + z)^2} \dr z r,
\end{eqnarray}
where
\begin{equation}\label{11.10}
B_2 \df \sqrt{\left(A_2 + 1\right)^2 - kr^2} - \sqrt{1 - kr^2}.
\end{equation}
Note that at the AH, where $A_1 = A_2 = B_2 = 0$, (\ref{11.9}) becomes $0 = 0$,
so expressions of the form $0/0$ will be present when integrating (\ref{11.9})
through the AH.

{}From (\ref{2.13}), using (\ref{11.4}) and (\ref{2.2}), we obtain
\begin{eqnarray}\label{11.11}
\dr z r &=& \frac {1 + z} {r \sqrt{1 - kr^2}} \left[\sqrt{\left(A_2 +
1\right)^2 - kr^2}\right. \nonumber \\
&+& \left.\frac {H_0 r (1 + z) \left(A_2 + 1\right)^2} {2 {\cal D}} \dr {t_B}
r\right].
\end{eqnarray}
Eliminating $\dril {t_B} r$ between (\ref{11.9}) and (\ref{11.11}) we get
\begin{equation}\label{11.12}
\dr z r = \frac {B_2 (1 + z)} {B_3 r \sqrt{1 - kr^2}} \left[\frac 3 2 - \frac
{kr^2} {\left(A_2 + 1\right)^2}\right],
\end{equation}
where
\begin{equation}\label{11.13}
B_3 \df \frac {A_1} {2 {\cal D}} + B_2 \frac {\sqrt{\left(A_2 + 1\right)^2 -
kr^2}} {\left(A_2 + 1\right)^2}.
\end{equation}
Using this in (\ref{11.11}) we get
\begin{eqnarray}\label{11.14}
\dr {t_B} r &=& \frac 1 {H_0 r (1 + z) B_3 \sqrt{\left(A_2 + 1\right)^2 -
kr^2}} \nonumber \\
&\times& \left\{{\cal D} {B_3} - A_1 \left[\frac 3 2 - \frac {kr^2} {\left(A_2 +
1\right)^2}\right]\right\}.\ \ \ \ \
\end{eqnarray}

In some of the numerical calculations, it will be more convenient to find $r(z)$
rather than $z(r)$, and for this purpose (\ref{11.12}) will be used in the form
\begin{equation}\label{11.15}
\dr r z = \frac {B_3 r \sqrt{1 - kr^2}} {B_2 (1 + z) \left[\frac 3 2 - \frac
{kr^2} {\left(A_2 + 1\right)^2}\right]}.
\end{equation}

\section{The limits of (\ref{11.12}) and (\ref{11.14}) at $r \to r_{\rm AH}$}
\label{limitsAHwithk}

\setcounter{equation}{0}

At the AH, where $A_1 = A_2 = 0$, we also have $B_2 = B_3 = 0$. Consequently, in
order to carry the integration through the AH in (\ref{11.12}) and
(\ref{11.14}), the expression $B_2 / B_3$ that becomes $0/0$ there must be
handled with care. This had already been noticed in Refs. \cite{KHBC2010} and
\cite{CBKr2010}, and Refs. \cite{CBKr2010} and \cite{McHe2008} demonstrated two
different solutions of this problem: in Ref. \cite{CBKr2010} an interpolating
polynomial, and in Ref. \cite{McHe2008} a Taylor expansion in $(z - z_{\rm AH})$
were used in place of the numerically calculated functions in the neighborhood
of the AH. In the case considered here, the limit of $B_2/B_3$ at the AH can be
explicitly calculated, as shown below.

{}From (\ref{11.12}) and (\ref{11.13}) we obtain
\begin{equation}\label{12.1}
\lim_{r \to r_{\rm AH}} \dr z r = \lim_{r \to r_{\rm AH}} \frac {(1 + z)
\left(3/2 - kr^2\right)} {r \sqrt{1 - kr^2} \left(\frac {A_1} {2 {\cal D} B_2}
+ \sqrt{1 - kr^2}\right)}.
\end{equation}
A simple calculation shows that
\begin{equation}\label{12.2}
\lim_{r \to r_{\rm AH}} \frac {A_1} {B_2} = \lim_{r \to r_{\rm AH}}
\left(\sqrt{1 - kr^2} \frac {A_1} {A_2}\right).
\end{equation}
Applying the de l'H\^{o}pital rule an making use of (\ref{9.2}) and of the fact
that $A_1 = 0$ on the AH, we find
\begin{equation}\label{12.3}
\lim_{r \to r_{\rm AH}} \frac {A_1} {A_2} = \lim_{r \to r_{\rm AH}}
\left(\Omega_m r {\cal D}^3 \dr z r\right).
\end{equation}
Let the following new symbol be introduced
\begin{eqnarray}\label{12.4}
{\cal G} &\df& \lim_{r \to r_{\rm AH}} \left[\left(1 - kr^2\right)^2 \right. \\
&+& \left.\left(1 - k r^2\right) \left(3 - 2 kr^2\right) \Omega_m {\cal D}^2 (1
+ z)\right]^{1/2}. \nonumber
\end{eqnarray}
Substituting (\ref{12.3}) and (\ref{12.2}) in (\ref{12.1}) and solving for
$\lim_{r \to r_{\rm AH}} \dril z r$ we obtain
\begin{equation}\label{12.5}
\lim_{r \to r_{\rm AH}} \dr z r = \lim_{r \to r_{\rm AH}} \frac {\left(3 -
2kr^2\right) (1 + z)} {r \left(1 - kr^2 + {\cal G}\right)}.
\end{equation}
Using this, $\lim_{r \to r_{\rm AH}} (B_2/B_3)$ can be easily calculated from
(\ref{11.12}), and the result used in (\ref{11.14}), to find
\begin{equation}\label{12.6}
\lim_{r \to r_{\rm AH}} \dr {t_B} r = \lim_{r \to r_{\rm AH}} \frac {2 {\cal D}
\sqrt{1 - kr^2}} {H_0 r (1 + z)} \left(\frac {3 - 2kr^2} {1 - kr^2 + {\cal G}} -
1\right).
\end{equation}
Equation (\ref{9.8}) is one more control value at the AH.

\section{The limits of (\ref{11.12}) and (\ref{11.14}) at $r \to 0$}
\label{limits0withk}

\setcounter{equation}{0}

Expressions of the form 0/0 also appear at $r = 0$. From (\ref{8.2}) one finds,
using $\Omega_m + \Omega_{\Lambda} \equiv 1$
\begin{equation}\label{13.1}
\lim_{r \to 0} \frac {\cal D} r = \lim_{r \to 0} \dr z r \df X.
\end{equation}
Anticipating that $X \neq 0$, so that $\lim_{r \to 0} \left(r^3 / {\cal
D}\right) = 0$, one finds from (\ref{9.2}), (\ref{9.5}), (\ref{11.10}) and
(\ref{11.13})
\begin{eqnarray}
\lim_{r \to 0} A_1 &=& \lim_{r \to 0} A_2 = \lim_{r \to 0} B_2 = -1,
\label{13.2} \\
\lim_{r \to 0} \frac r {A_2 + 1} &=& \sqrt{\frac X {2M_0 H_0}}, \label{13.3} \\
\lim_{r \to 0} \left(rB_3\right) &=& - \frac 1 {2X} - \sqrt{\frac X {2M_0 H_0}}\
\sqrt{1 -
\frac {k X} {2M_0 H_0}}. \nonumber \\
\label{13.4}
\end{eqnarray}
Taking the limit of (\ref{11.12}) at $r \to 0$, then using (\ref{13.2}) --
(\ref{13.4}), we obtain
\begin{equation}\label{13.5}
X^3 + kX - 2M_0 H_0 = 0.
\end{equation}
This equation, irrespectively of the value of $k$, has only one solution such
that $X > 0$; a proof is given in Appendix \ref{oneX}.\footnote{$\lim_{r \to 0}
\dril z r < 0$ would mean that $z < 0$ in a vicinity of the observer, i.e., that
the Universe locally collapses upon her. While this happens in certain
inhomogeneous models, this cannot happen in a model designed to mimic RW.} This
solution is located in $(U_1, U_2)$, where
\begin{eqnarray}\label{13.6}
U_1 &=& \left(2M_0 H_0\right)^{1/3}, \\
U_2 &>& \sqrt{- k/3} + {\rm max} \left\{\left(2M_0 H_0\right)^{1/3}, \sqrt{-
2k}/3\right\}. \nonumber
\end{eqnarray}
Using (\ref{13.5}) back in (\ref{13.4}) we obtain
\begin{equation}\label{13.7}
\lim_{r \to 0} \left(rB_3\right) = - \frac 3 {2X} + \frac k {2M_0 H0}.
\end{equation}

{}From (\ref{8.1}) and (\ref{13.1}) we have
\begin{equation}\label{13.8}
\lim_{r \to 0} \frac {R_{\rm ng}} r = \lim_{r \to 0} \frac {\cal D} {H_0 r (1 +
z)} = \frac X {H_0},
\end{equation}
and then from (\ref{11.5}) -- (\ref{11.6})
\begin{eqnarray}\label{13.9}
&& {\cal T} \df t(0) - t_B(0) \\
&& = \frac {M_0} {(-k)^{3/2}} \left[\sqrt{{\cal Y}^2 - 1} - \ln \left({\cal Y} +
\sqrt{{\cal Y}^2 - 1}\right)\right], \nonumber
\end{eqnarray}
where
\begin{equation}\label{13.10}
{\cal Y} \df 1 - \frac {kX} {M_0 H_0}.
\end{equation}
The ${\cal T}$ in (\ref{13.9}) is the age of the Universe at $r = 0$.

{}From (\ref{11.14}) we can further calculate
\begin{eqnarray}\label{13.11}
\lim_{r \to 0} \dr {t_B} r &=& \frac {M_0 X} {2 \left(3 M_0 H_0 - kX\right)} \\
&\times& \left[3 \left(\Omega_m - 1\right) - \frac {kX} {M_0 H_0} \left(\frac 3
2\ \Omega_m - 1\right)\right]. \nonumber
\end{eqnarray}
This calculation is tricky, so it is presented in Appendix \ref{limof1311}. With
$k < 0$ and $\Omega_m = 0.32$, the limit (\ref{13.11}) is negative, so $t_B(r)$
will be a decreasing function of $r$ at least in some neighbourhood of $r = 0$.

\section{The age of the Universe and the value of $k$}\label{ageandk}

\setcounter{equation}{0}

The numerical values of the constants that will appear in the calculations are
given by (\ref{10.1}).

Before proceeding to solve (\ref{11.12}), a value for $k$ must be chosen. That
value determines the age of the Universe at the center, via (\ref{13.9}) and
(\ref{13.10}). It is to be noted that $X$, given by (\ref{13.5}), is a function
of $k$. For $k < 0$, $X > \left(2 M_0 H_0\right)^{1/3}$ must hold, and $X$
monotonically increases from $\left(2 M_0 H_0\right)^{1/3}$ at $k = 0$ to $+
\infty$ at $k \to - \infty$ (but $\dril X k \llim{k \to - \infty} 0$).

\begin{figure}
\hspace{-4.5cm}
\includegraphics[scale=0.55]{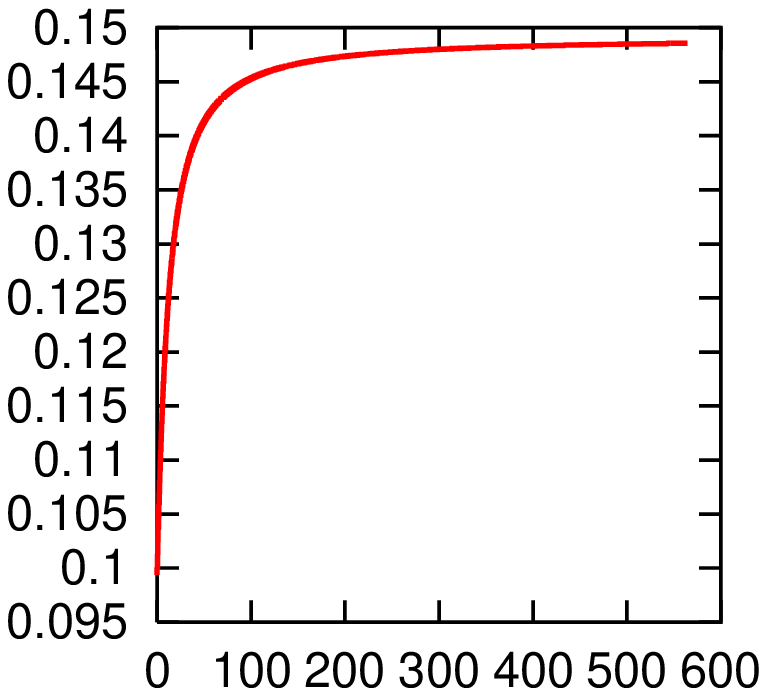}
 ${}$ \\[-4cm]
\hspace{4cm}
\includegraphics[scale = 0.5]{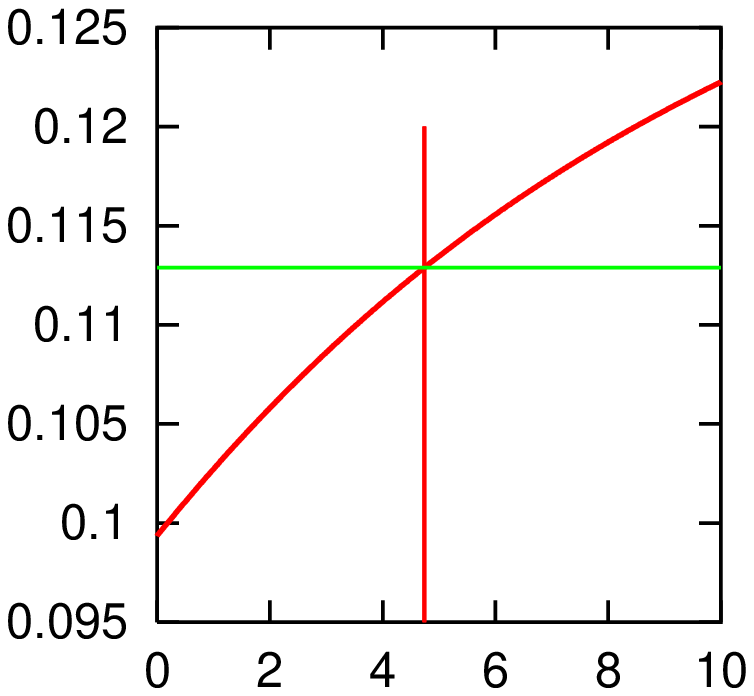}
\caption{{\bf Left panel:} Graph of the function ${\cal T}(-k)$. It has the
upper bound $1/H_0$. {\bf Right panel:} A closeup view of the same graph over a
smaller range of $k$. The vertical line marks the value $k = - 4.7410812$ chosen
in numerical experiments (see text), the horizontal line marks the associated
age of the Universe at the center ${\cal T}(0) = 0.1128971437689653$ NTU.}
\label{drawageoftau}
\end{figure}

For the function ${\cal T}(k)$ given by (\ref{13.9}) we find
\begin{eqnarray}
\lim_{k \to 0} {\cal T} &=& \frac 2 {3 H_0}, \qquad \lim_{k \to - \infty} {\cal
T} = \frac 1 {H_0} \label{14.1} \\
\lim_{k \to - \infty} \dr {\cal T} k &=& 0. \label{14.2}
\end{eqnarray}
The graph of ${\cal T}(-k)$ is shown in Fig. \ref{drawageoftau}. As can be seen,
${\cal T}(-k) < 1/H_0$ everywhere. However, in the L--T model, the ``age of the
Universe'' is different at every $r$. The point, at which the L--T age can be
compared to that of $\Lambda$CDM is the intersection of the past light cone of
the L--T observer with the Big Bang. This is the place that the observer can (in
principle) see and infer something about, not the central age given by
(\ref{13.9}) -- (\ref{13.10}). The L--T age at that point (call it ``edge age'')
could be assumed equal to (\ref{10.9}), and the corresponding value of $k$ could
then be determined in numerical experiments. This can be done by assuming a
value for ${\cal T}$, calculating $k$ from (\ref{13.9}) -- (\ref{13.10}), then
solving (\ref{11.12}) -- (\ref{11.15}), and deducing a correction to the chosen
value of ${\cal T}$.

But it turns out that a preferred unique value of $k$ emerges already by
integrating (\ref{11.12}) (or (\ref{11.15})). With $k$ off the preferred value,
the function $z(r)$ found by integrating (\ref{11.15}) backward from $r = r_{\rm
AH}$, misses the point $(r, z) = (0, 0)$. The $k$ fine-tuned to ensure that
$z(0) = 0$ implies the edge age close to (\ref{10.9}).

However, there is a problem here. Similarly to what has been said above, when $k$
is off the opitmal value, $z(r)$ found by integrating (\ref{11.12}) forward from
$r = 0$, misses the point $(r, z) = (r, z)_{\rm AH}$. The $k$ that ensures
maximal precision at $r = 0$ is $- 4.74061$, the one that ensures maximal
precision at $r_{\rm AH}$ is $- 4.7410812$. A preference was given to maximal
precision at $r_{\rm AH}$. So, the $k$ fine-tuned to $z_{\rm AH}$ and its
associated $X$ (found from (\ref{13.5}) by the bisection method)
are\footnote{These numbers, and several other numbers displayed further on, may
look to be excessively precise. They are indeed -- for astrophysical
applications. However, this precision is necessary to avoid misalignments in some
of the graphs. It may also be necessary for those other authors who might wish to
verify and reproduce the results presented here.}
\begin{equation}\label{14.3}
k = - 4.7410812, \qquad X = 3.028567231968699.
\end{equation}

The reason why the value of $k$ is determined already by (\ref{11.2}) (with
$z(0)$ and $z(r_{\rm AH})$ given) is the following. Equation (\ref{11.2}) that
determines $z(r)$ is of first order, so its solution is uniquely determined by
$z(0)$ or $z(r_{\rm AH})$ alone. If both $z(0)$ and $z(r_{\rm AH})$ are
specified, then a limitation is imposed on the solution that determines the value
of the single free parameter in $z(r)$, which is $k$. It follows that an $E = 0$
L--T model cannot obey (\ref{8.1}).

\section{Numerical calculation of $z(r)$ from
(\ref{11.12})}\label{numericszwithk}

\setcounter{equation}{0}

The precision in calculating $z(r)$ and $t_B(r)$ depends on the precision in
determination of the function ${\cal D}(z)$ and of the values of ${\cal D}$ and
$z$ at the AH. So, first, a Fortran 90 program was used to determine ${\cal
D}(z)$ for any $z$ by calculating the integral in (\ref{8.2}) with the step
${\rm d} z' = z_{\rm max} \times 10^{-9}$ (the same program that produced the
data for Fig. \ref{drawdfun}, but more precise) up to $z_{\rm max} = 1.585$ --
slightly above the $z_{\rm AH}$ from Fig. \ref{drawdfun}. This program found the
values of $z_{\rm AH}$ and ${\cal D}_{\rm AH}$ (at which $A_1 = 0$) to be
\begin{eqnarray}\label{15.1}
&& \hspace{-1.5cm} 1.582432259768032 < z_{\rm AH} < 1.582432261353032,
\nonumber \\
&& \hspace{-1.5cm} 1.037876550094136 < {\cal D}_{\rm AH} < 1.037876550731146.
\end{eqnarray}
These lower limits of $z_{\rm AH}$ and ${\cal D}_{\rm AH}$ were provisionally
taken as their true values. The interval $Z \df [0, z_{\rm AH}]$ was divided
into $10^5$ segments, for each $z_i \in Z$, $i = 0, \dots, 10^5 - 1$ the value
of ${\cal D}(z_i)$ was found, and the $(z_i, {\cal D}_i)$ were tabulated.
Numerical errors caused that the last value of $z$ in the table was larger than
the upper limit in (\ref{15.1}). Consequently, the penultimate values of $z$ and
${\cal D}$ were taken as defining the AH, they are
\begin{equation}\label{15.2}
(z, {\cal D})_{\rm AH} = (1.582430687623614, 1.037876401742206),
\end{equation}
and they are both lower than the lower limits in (\ref{15.1}). The corresponding
$r_{\rm AH}$ was calculated from (\ref{9.4}):
\begin{equation}\label{15.3}
r_{\rm AH} = 0.3105427968086945.
\end{equation}
The table of values of ${\cal D}(z)$ was then used in integrating (\ref{11.15})
numerically backward from $z = z_{\rm AH}$ to $z = 0$.

\begin{figure}[h]
 ${}$ \\[0.5cm]
\hspace{-0.5cm}
\includegraphics[scale = 0.85]{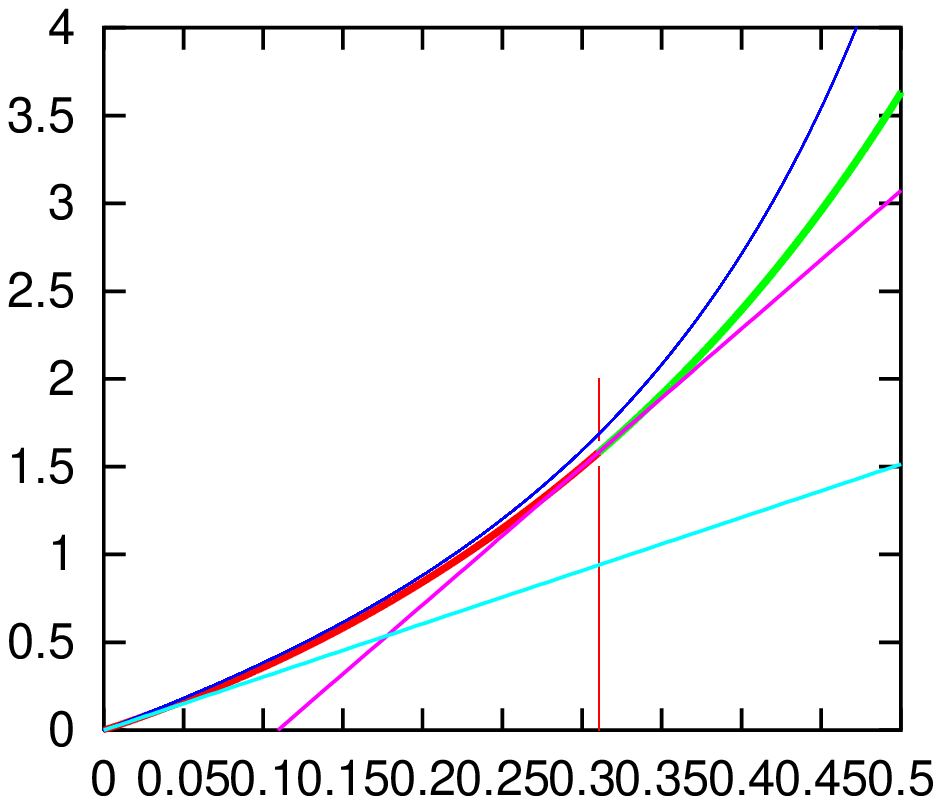}
 ${}$ \\[-6.7cm]
\hspace{-2cm}
\includegraphics[scale = 0.4]{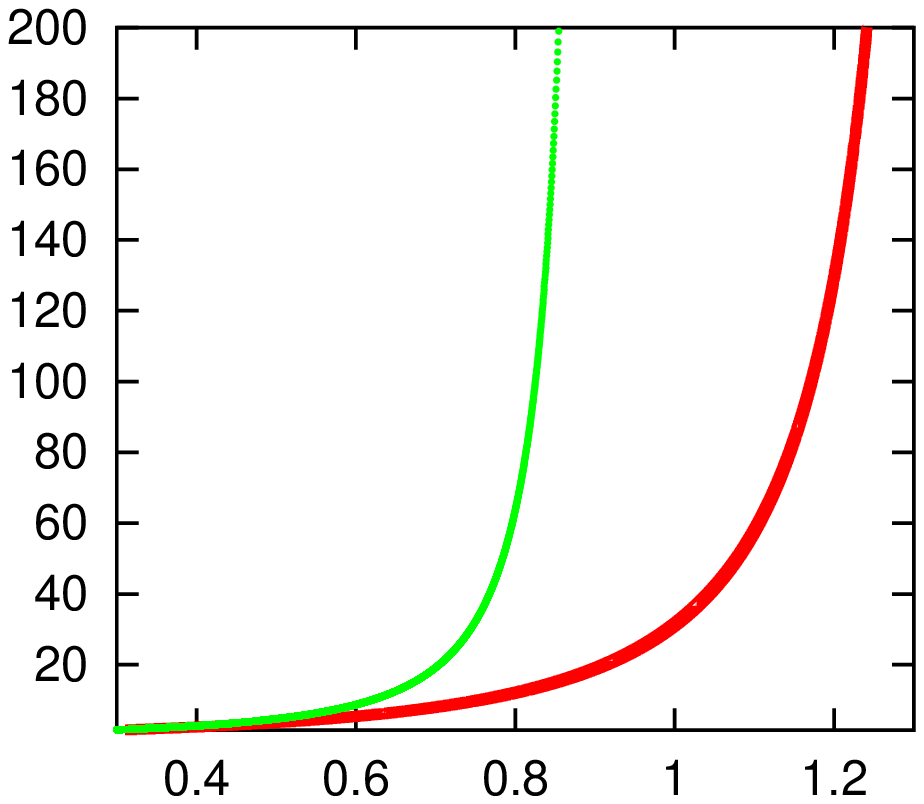}
 ${}$ \\[3.5cm]
\includegraphics[scale = 0.4]{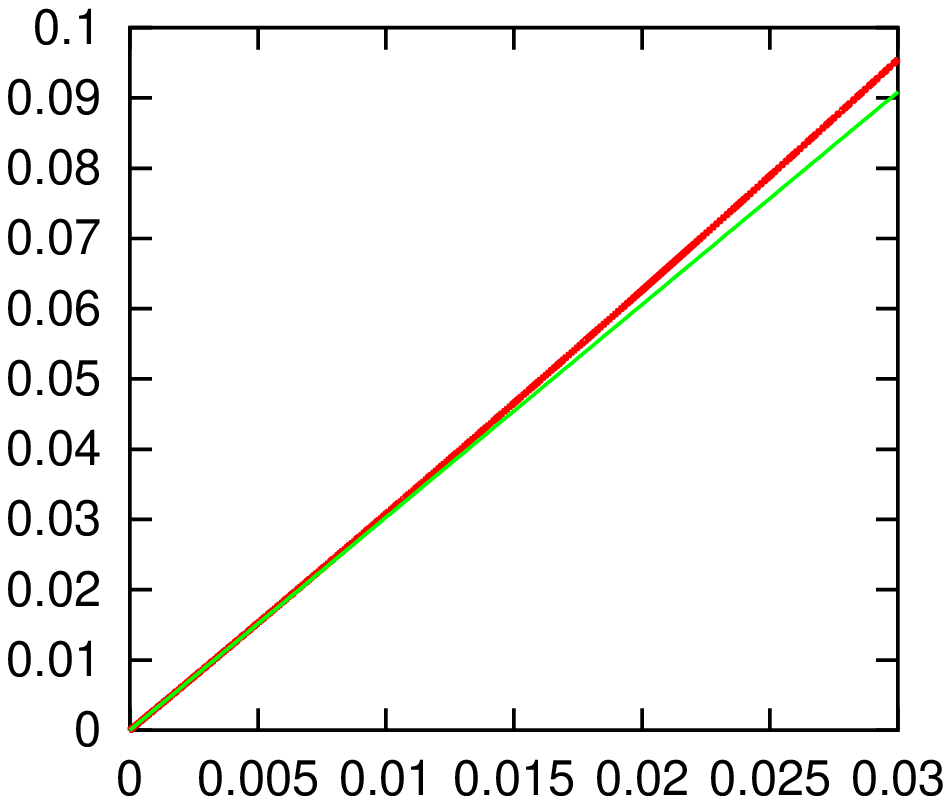}
\includegraphics[scale = 0.4]{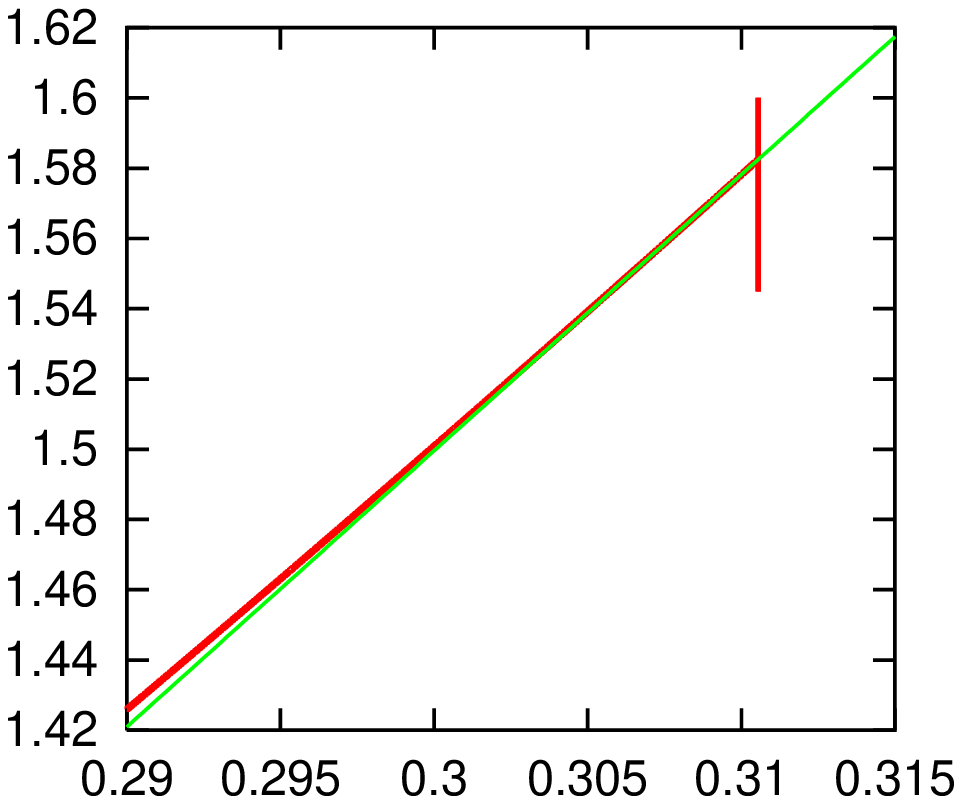}
\caption{The lower curve in the large panel is the graph of $z(r)$ calculated by
integration of (\ref{11.15}) backward and of (\ref{11.12}) forward from the AH,
with $k = -4.7410812$. The upper curve is the function $z_{\Lambda}(r)$ of the
$\Lambda$CDM model. The vertical line marks $r = r_{\rm AH}$ given by
(\ref{15.2}). The sloping straight lines are tangents to $z(r)$ at $r = 0$ and
at $r = r_{\rm AH}$ calculated from (\ref{13.5}) and (\ref{12.5}), respectively.
The right curve in the inset is $z(r)$ for $r > r_{\rm AH}$. The left curve is
the corresponding segment of $z_{\Lambda}(r)$. The two lower panels show closeup
views of the neighbourhood of $r = 0$ (left) and of $r = r_{\rm AH}$ (right).
The vertical stroke in the right panel marks $r = r_{\rm AH}$.}
\label{drawznointerwithk}
\end{figure}

Since $z \to \infty$ at the Big Bang, $z$ is not a usable parameter in the
vicinity thereof, and ${\cal D}(z)$ cannot be tabulated in that region. The
value of $r$, at which the Big Bang would be reached, was not known in advance,
so it took some experimenting to determine the step $\Delta r = 2.4 r_{\rm AH}
\times 10^{-5}$ and the number of steps $N = 15 \times 10^4$. For each $r >
r_{\rm AH}$, the corresponding $z$ was calculated by integrating (\ref{11.12})
forward, with the initial condition (\ref{15.2}), and the corresponding ${\cal
D}(z)$ was found from
\begin{equation}\label{15.4}
{\cal D}(z + \Delta z) = {\cal D}(z) + \frac {\Delta z} {\sqrt{\Omega_m (1 +
z')^3 + \Omega_{\Lambda}}},
\end{equation}
which is a consequence of (\ref{8.2}). The biggest values of $(r, z)$ that the
program could yet handle were
\begin{eqnarray}\label{15.5}
&& (r, z)_{\rm BB} = \left(1.422005301219788\right., \nonumber \\
&&\ \ \ \ \ \ \ \ \  \left.1.6236973619875722 \times 10^{229}\right).
\end{eqnarray}
The $r_{\rm BB}$ was taken to be at the intersection of the observer's past
light cone with the Big Bang.

The resulting function $z(r)$ is presented in Fig. \ref{drawznointerwithk}; for
later reference it will be denoted by $z_{\rm back}(r)$. The main graph shows
$z(r)$ for $r \in [0.0, 0.5]$ (the lower curve). The upper curve is the function
$z_{\Lambda}(r)$ of the $\Lambda$CDM model, calculated from (\ref{6.1}),
(\ref{4.4}) and (\ref{2.16}). The right curve in the inset is $z(r)$ for $r \in
[0.3, 1.3]$, i.e. from the neighbourhood of the AH to a value at which $z$
begins to grow very fast. The left curve is $z_{\Lambda}(r)$ in the same range
of $r$. The panels below the main graph show that $z(r)$ respects the slopes
given by (\ref{13.5}) and (\ref{12.5}) at $r = 0$ and $r = r_{\rm AH}$,
respectively, with a satisfactory precision.

As seen from Fig. \ref{drawznointerwithk}, the functions $z(r)$ in the L--T
model and in the $\Lambda$CDM model are different. In particular,
$z_{\Lambda}(r) \to \infty$ at $r = 0.9098426708844661 < r_{\rm BB}$. Thus, this
time it should not be expected that the light cone of the L--T model will
coincide with that of $\Lambda$CDM. The aim here is not to duplicate the light
cone, but the $D_L(z)$ relation (\ref{2.20}) via (\ref{2.14}).

In order to verify the precision of the algorithm, the calculation of $z(r)$ was
repeated by a different method. Namely, (\ref{11.12}) was integrated from $r =
0$ up to a point close behind the AH. For each $z$, the associated value of
${\cal D}$ was calculated from (\ref{15.4}). The number of steps was $11 \times
10^4$, and the size of the step was $\Delta r = 10^{-5}r_{\rm AH}$.

A problem occurred near $r = r_{\rm AH}$. Namely, because of numerical errors,
$A_2$ became zero at a smaller $r$ than $A_1$, even though each of them is
supposed to become zero at $r = r_{\rm AH}$. As a result, in the range where
$A_2 > 0$ (and thus $B_2 > 0$) while $A_1 < 0$, $\dril z r$ calculated by the
program became negative and could not return to positive values when the
calculation was continued. Thus, the program was designed to stop once $A_2$
becomes positive. The function thus obtained will be denoted $z_{\rm forw}(r)$.

\begin{figure}[h]
${ }$ \\[0.2cm]
\includegraphics[scale = 0.4]{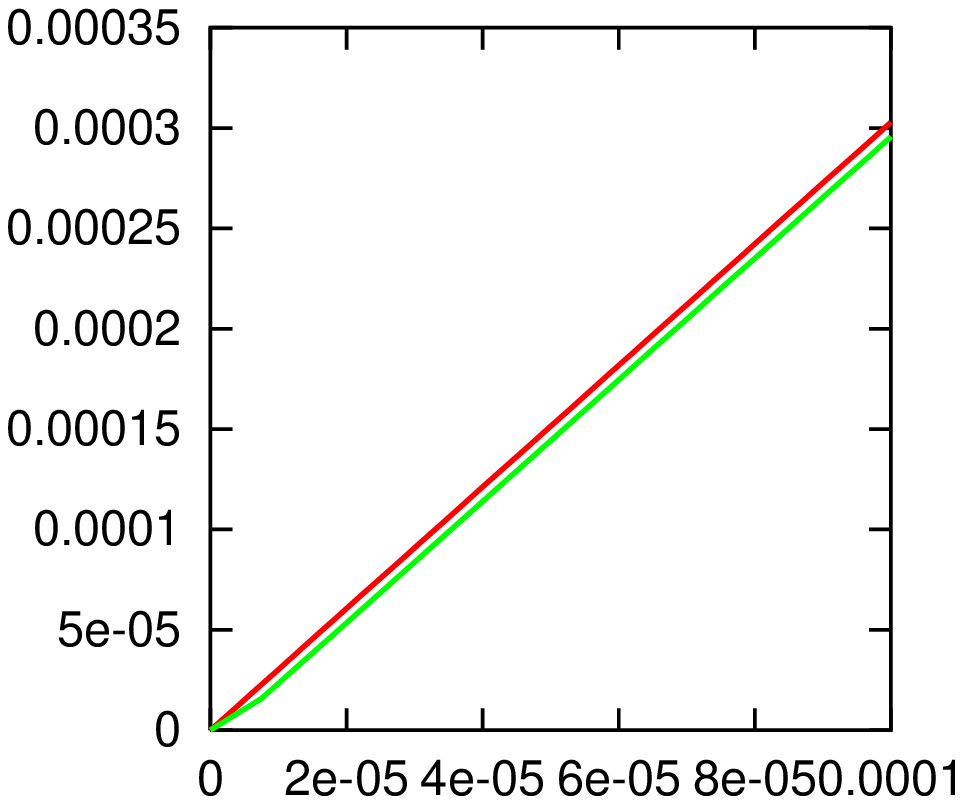}
\includegraphics[scale = 0.4]{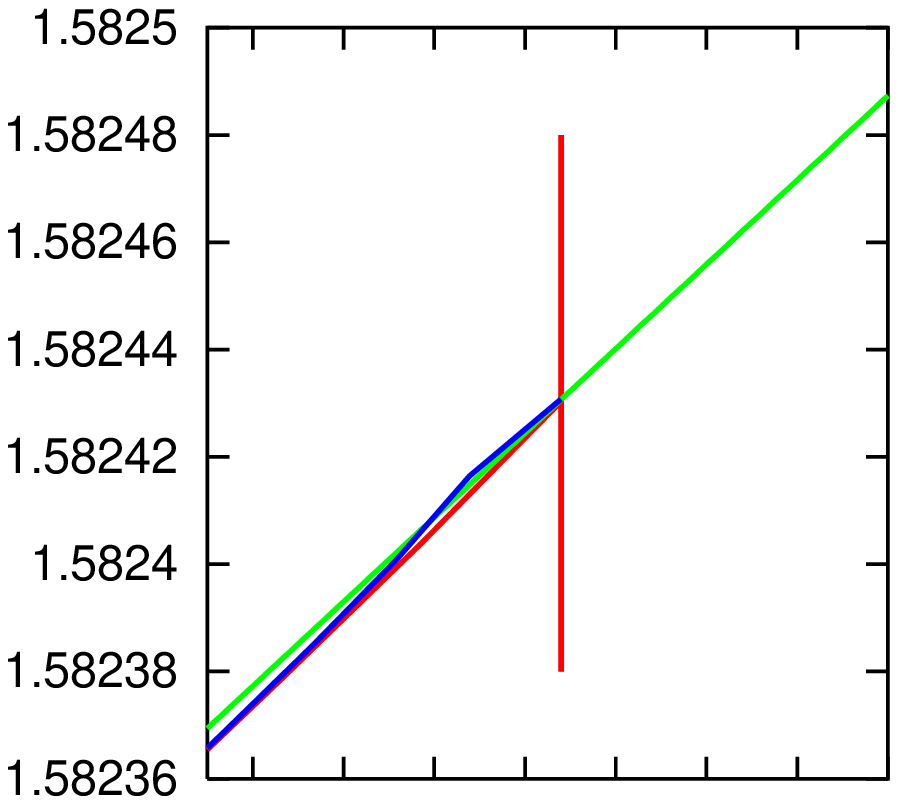}
\caption{Comparison of $z_{\rm back}(r)$ and $z_{\rm forw}(r)$. {\bf Left
panel:} closeup view of the segment $r \in [0,10^{-4}]$. The upper line is
$z_{\rm forw}(r)$; the lower line is $z_{\rm back}(r)$. {\bf Right panel:}
closeup view of the vicinity of $r = r_{\rm AH}$ ($r_{\rm AH}$ is marked with
the vertical stroke). The sloping straight line is the tangent given by
(\ref{12.5}). The broken line is $z_{\rm back}(r)$. The third line is $z_{\rm
forw}(r)$; at the scale of this figure it seems to hit the end point $(r_{\rm
AH}, z_{\rm AH})$ exactly. The tics on the horizontal axis go from $0.310536$ to
$0.310550$ and are separated by $\Delta r = 2 \times 10^{-6}$.}
\label{drawzfrom0nointerwithk}
\end{figure}

At the scale of Fig. \ref{drawznointerwithk}, the graphs of $z_{\rm forw}(r)$
and $z_{\rm back}(r)$ coincide. Near $r = 0$ they differ by $\Delta z \approx
7.35 \times 10^{-4}$; see the left panel of Fig. \ref{drawzfrom0nointerwithk}.
As explained in Sec. \ref{ageandk}, the precision in that area could be improved
to $0.5 \times 10^{-7}$, but this would cause a worse precision at the AH, where
the difference between the two curves is $\Delta z \approx 0.5 \times 10^{-6}$.
The right panel shows that area; at that scale the end points of the two curves
seem to coincide.

\section{Numerical calculation of $t_B(r)$ from
(\ref{11.14})}\label{numtB}

\setcounter{equation}{0}

Several quantities in (\ref{11.14}) tend to zero as $r \to 0$. It is important
not to let the numerical program divide a finite quantity by one that tends to
zero. Consequently, it is advantageous to rearrange (\ref{11.14}) by introducing
new quantities as follows, using (\ref{11.13}) and (\ref{11.9}):
\begin{eqnarray}
F_1 &\df& {\cal D} / r \llim{r \to 0} X, \label{16.1} \\
F_2 &\df& \frac {F_1} {2M_0 H_0 (1 + z)}, \label{16.2} \\
F_3 &\df& \sqrt{1 - kF_2}, \label{16.3} \\
rB_3 &\df& F_4 = \frac {A_1} {2 F_1} + B_2 \sqrt{F_2} F_3, \label{16.4} \\
F_5 &\df& \frac {\sqrt{F_2}} {H_0 (1 + z) F_3 F_4}, \label{16.5}
\end{eqnarray}
 \vspace{-0.4cm}
\begin{eqnarray}
&& {\cal D} B_3 - A_1 \left[\frac 3 2 - \frac {kr^2} {\left(A_2 +
1\right)^2}\right] \df F_6 \label{16.6} \\
&&= F_3 \left[- F_1 \sqrt{F_2} \sqrt{1 - kr^2} + F_3 \frac {1 + z}
{\sqrt{\Omega_m (1 + z)^3 + \Omega_{\Lambda}}}\right], \nonumber
\end{eqnarray}
\begin{equation}
\dr {t_B} r = F_3 F_5 \left(\frac {F_6} r\right). \label{16.7}
\end{equation}
Of the quantities defined above, $F_1$ and $(F_6/r)$ behave as 0/0 at $r = 0$.
However, $\lim_{r \to 0} F_1$ is given by (\ref{13.1}) and (\ref{13.5}), and the
value of $F_1$ at the first grid point after $r = 0$ is calculated without
problems using (\ref{11.12}). Given $F_1$, the values of $F_2, \dots, F_5$ at $r
= 0$ are well-defined, and the only remaining 0/0 expression is $F_6/r$.

The parametrisation (\ref{16.1}) -- (\ref{16.6}) works well in a neighbourhood
of $r = 0$. At $r = r_{\rm AH}$ other quantities in (\ref{11.4}) tend to zero
(they are $A_1$, $A_2$, $B_2$ and $B_3$), and another rearrangement minimises
numerical errors:
\begin{eqnarray}
G_1 &\df& A_2/A_1, \label{16.8} \\
G_2 &\df& \sqrt{\left(A_2 + 1\right)^2 - kr^2}, \label{16.9} \\
G_3 &\df& \sqrt{1 - kr^2}, \label{16.10} \\
G_4 &\df& H_0 (1 + z) G_2, \label{16.11} \\
\frac {B_3} {A_1} \df G_5 &=& \frac 1 {2 {\cal D}} + \frac {G_1 G_2 \left(A_2 +
2\right)} {\left(A_2 + 1\right)^2 \left(G_2 + G_3\right)},\ \ \ \  \label{16.12}
\end{eqnarray}
 \vspace{-0.4cm}
\begin{equation}\label{16.13}
\dr {t_B} r = \frac 1 {rG_4}\ \left\{{\cal D} - \frac 1 {G_5} \left[\frac 3 2 -
\frac {kr^2} {\left(A_2 + 1\right)^2}\right]\right\}.
\end{equation}
A similar rearrangement must be made in (\ref{11.12}),
\begin{eqnarray}
\frac {B_2} {A_1} \df G_6 &=& \frac {A_2 + 2} {G_2 + G_3}\ G_1, \label{16.14} \\
\frac {r B_3} {B_2} \df G_7 &=& \frac r {2 {\cal D} G_6} + \frac {r G_2}
{\left(A_2 + 1\right)^2}, \label{16.15} \\
\dr z r &=& \frac {1 + z} {G_3 G_7}\ \left[\frac 3 2 - \frac {k r^2} {\left(A_2
+ 1\right)^2}\right].\ \ \  \label{16.16}
\end{eqnarray}

The only quantity in (\ref{16.8}) -- (\ref{16.16}) that behaves like 0/0 at $r
\to r_{\rm AH}$ is $G_1$. The $G_2$, $G_3$ and $G_4$ have well-defined values at
$r_{\rm AH}$, and once $G_1$ is calculated, $G_5$, $G_6$ and $G_7$ have values
at $r_{\rm AH}$, too. Experiment showed that the program handles $G_1$ without
any fluctuations.

It would be natural to combine the two rearrangements so that as many
occurrences of $r$ as possible in (\ref{16.8}) -- (\ref{16.16}) cancel out, thus
hopefully improving the accuracy of this parametrisation near $r = 0$. Such an
experiment was done, but it did not lead to the intended improvement -- the
graph of $t_B(r)$ did not change.

The limit of $(\dril {t_B} r)$ at $r \to r_{\rm AH}$ is given by (\ref{12.6}),
but $t_B(r_{\rm AH})$ cannot be calculated independently of (\ref{11.14}). The
integration of (\ref{16.7}) must thus begin at $r = 0$, and $t_B(r_{\rm AH})$ is
found in the process. The $t_B(0)$ is given by (\ref{13.9}), with $t(0) = 0$ by
assumption.

\begin{figure}[h]
\hspace{-0.7cm}
\includegraphics[scale = 0.85]{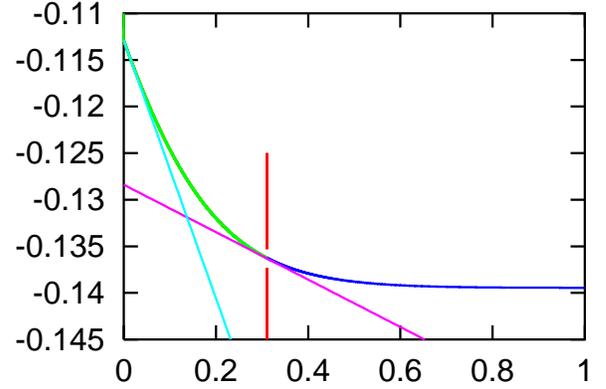}
\caption{The function $t_B(r)$ calculated by integrating (\ref{16.7}) from $r =
0$ to $r = r_{\rm AH}$ and by integrating (\ref{16.13}) beyond $r = r_{\rm AH}$.
The stroke marks $r = r_{\rm AH}$. The sloping straight lines are the tangents
to $t_B(r)$ at $r = 0$ and at $r = r_{\rm AH}$ calculated from (\ref{13.11}) and
(\ref{12.6}), respectively. The curve found by integrating (\ref{16.13})
backward from $r = r_{\rm AH}$ is also present in the figure, but, at this
scale, it coincides with the $r < r_{\rm AH}$ part of first curve. The
discrepancies between the two curves are shown in Figs.
\ref{drawtbfrom0withkat0} and \ref{drawtbfrom0withkatAH}. }
\label{drawtbfrom0withk}
\end{figure}

Equation (\ref{16.7}) was integrated from $r = 0$ to $r = r_{\rm AH}$ by using
the tabulated values of $z$ and ${\cal D}$, with $r(z)$ calculated along the way
from (\ref{11.15}). A continuation of $t_B(r)$ for $r > r_{\rm AH}$ was found by
integrating (\ref{16.13}) and calculating the values of $z$ and ${\cal D}$ from
(\ref{11.12}) and (\ref{15.4}), respectively. The initial point for the
continuation was corrected as described further on. Figure
\ref{drawtbfrom0withk} shows the resulting $t_B(r)$, together with the tangents
at $r = 0$ and $r = r_{\rm AH}$, and with the vertical line marking $r = r_{\rm
AH}$.

\begin{figure}[h]
\hspace{-0.7cm}
\includegraphics[scale = 0.85]{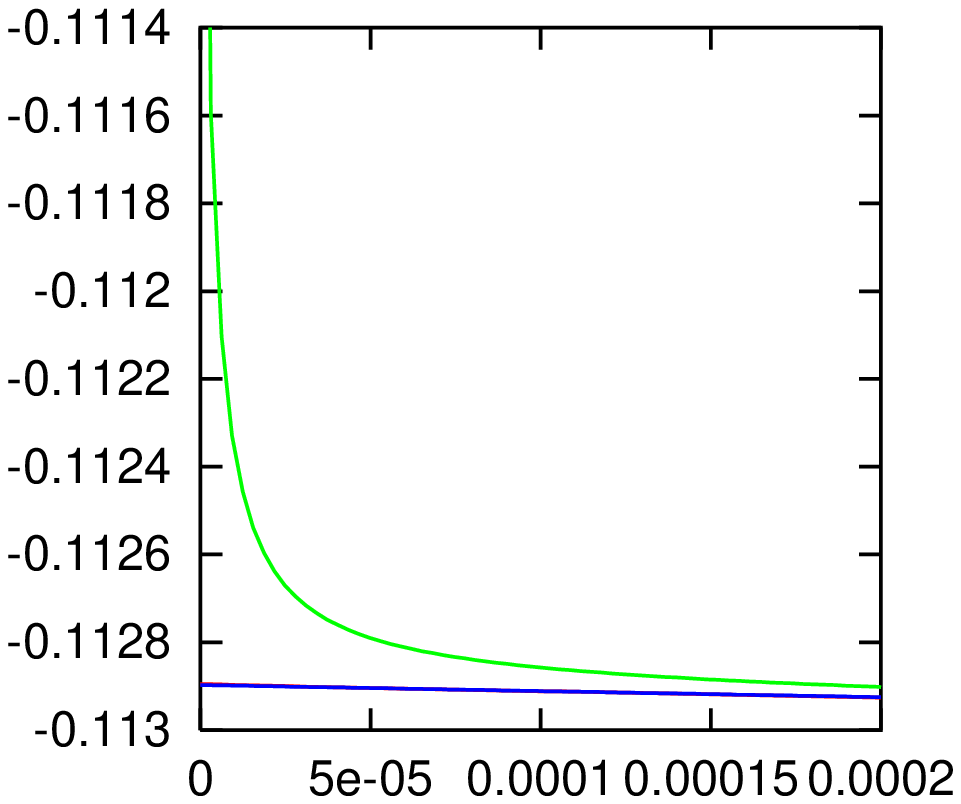}
 ${}$ \\[-6.5cm]
\hspace{1cm}
\includegraphics[scale = 0.5]{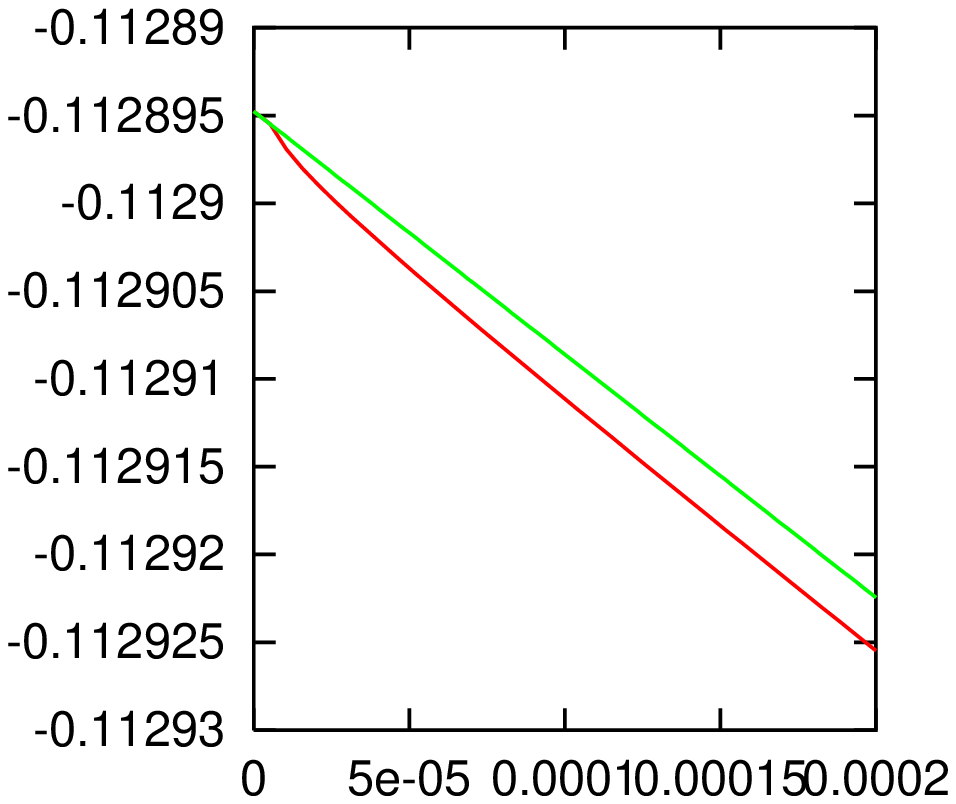}
\vspace{2.5cm} \caption{Closeup view of the neighbourhood of $r = 0$ in Fig.
\ref{drawtbfrom0withk}. The upper curve in the main graph is $t_B(r)$ calculated
by integrating (\ref{16.13}) backward from $r = r_{\rm AH}$. The lower curve is
obtained by integrating (\ref{16.7}) forward from $r = 0$; at this scale, within
the figure, it coincides with its tangent given by (\ref{13.11}). The inset
shows that the lower curve departs from the correct slope already at the first
grid point, but this leads to the difference $\Delta t_B < 0.3 \times 10^{-7}$
NTU $\approx 2940$ years.} \label{drawtbfrom0withkat0}
\end{figure}

\begin{figure}[h]
\hspace{-0.7cm}
\includegraphics[scale = 0.85]{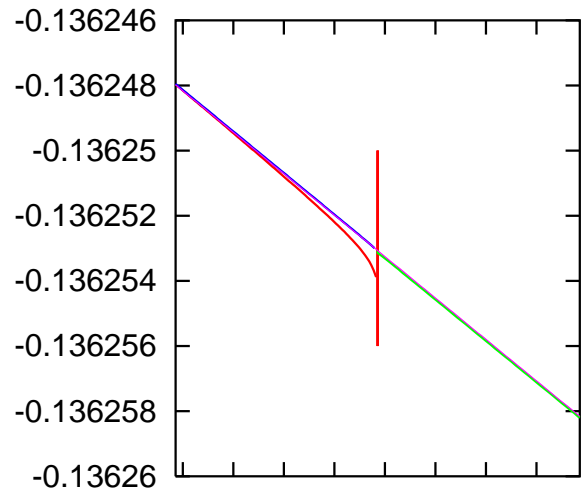}
\caption{Closeup view of the curves from Fig. \ref{drawtbfrom0withk} in the
neighbourhood of $r = r_{\rm AH}$ (marked by the vertical stroke). The sloping
straight line is the tangent from (\ref{12.6}). The $t_B(r)$ curves, calculated
by integrating (\ref{16.13}) forward and backward from $r = r_{\rm AH}$,
coincide at this scale with the tangent. Their initial point had to be set by
hand as described in the text. The lower curve left of $r_{\rm AH}$ is the
$t_B(r)$ found by integrating (\ref{16.7}) forward from $r = 0$. It misses the
slope at $r = r_{\rm AH}$ given by (\ref{12.6}). The tics on the horizontal axis
go from $0.31035$ to $0.3107$ with the interval $\Delta r = 5 \times 10^{-5}$. }
\label{drawtbfrom0withkatAH}
\end{figure}

In order to verify the calculation of $t_B(r)$, (\ref{16.13}) was integrated
backward from $r = r_{\rm AH}$, with the initial value $t_B(r_{\rm AH})$
corrected as described below. The resulting curve is also shown in Fig.
\ref{drawtbfrom0withk}, but, at this scale, looks to coincide with the former
one. Figures \ref{drawtbfrom0withkat0} and \ref{drawtbfrom0withkatAH} display
the discrepancies between the two integrations. As seen in the main graph of
Fig. \ref{drawtbfrom0withkat0}, the integration backward from $r_{\rm AH}$ gives
a large discrepancy with the initial data at $r = 0$ given by (\ref{13.9}) and
(\ref{13.11}). At $r \approx 0.002$ the difference is $\Delta t_B \approx 2.0
\times 10^{-5}$ NTU $\approx 1.96 \times 10^6$ years. The inset in Fig.
\ref{drawtbfrom0withkat0} displays an even closer look at the neighbourhood of
$r = 0$. It shows the forward-integrated $t_B(r)$ (the lower curve) and its
tangent calculated from (\ref{13.11}). Numerical instabilities cause that the
curve departs from the right slope already at the first grid point, but the
resulting difference $\Delta t_B < 0.3 \times 10^{-7}$ NTU $\approx 2940$ years,
which is cosmologically insignificant.

Figure \ref{drawtbfrom0withkatAH} shows the neighbourhood of $r = r_{\rm AH}$ in
Fig. \ref{drawtbfrom0withk}. The $t_B(r)$ found by integrating (\ref{16.7})
forward from $r = 0$ is the lower curve left of the vertical stroke, which marks
$r = r_{\rm AH}$. It misses the correct slope at $r_{\rm AH}$ -- the tangent at
$r_{\rm AH}$, calculated from (\ref{12.6}), is the sloping straight line. (This
means it also missed the correct value at $r_{\rm AH}$, but this cannot be
calculated independently.) However, it coincides with that tangent in a certain
range of $r$, so the intersection of the tangent with $r = r_{\rm AH}$ at $t =
-0.1362530696173036$ was assumed to be the correct end point $t_B(r_{\rm AH})$.
This end point was then used as the initial point for the integration of
(\ref{16.13}) forward and backward from $r = r_{\rm AH}$. Both integrations
avoided numerical instabilities. At the scale of Fig.
\ref{drawtbfrom0withkatAH}, these curves coincide with the tangent.

The graphs indicate that $t_B(r)$ is a decreasing function in the whole range,
so no shell crossings are present.

\section{Numerical calculation of the light cone}\label{numcone}

\setcounter{equation}{0}

With $t_B(r)$ now given, eq. (\ref{11.2}) can be numerically solved.
Substituting (\ref{11.4}) in (\ref{11.2}), using (\ref{8.1}) -- (\ref{8.2}) for
$R$, and then using (\ref{16.1}) -- (\ref{16.3}) we obtain
\begin{equation}\label{17.1}
\dr t r = \frac 1 {\sqrt{1 - kr^2}} \left[- \frac {F_1} {H_0 (1 + z)}
+ r \dr {t_B} r \frac {F_3} {\sqrt{F_2}}\right].
\end{equation}
This is well-behaved at $r = 0$ and at $r = r_{\rm AH}$. The values of
$t_{B,r}(r)$ were found in integrating (\ref{16.7}) and (\ref{16.13}).

\begin{figure}[h]
\hspace{-0.7cm}
\includegraphics[scale=0.85]{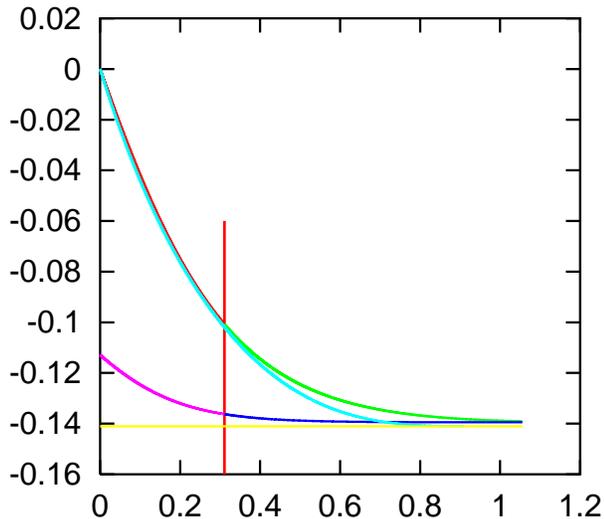}
\caption{The past light cone of the central observer in the L--T model that
duplicates the relation (\ref{8.1}) (the uppermost curve) compared with that of
the $\Lambda$CDM model (\ref{4.4}) (the lower curve, partly nearly coincident
with the first one). The lowest curve is the $t_B(r)$ from Fig.
\ref{drawtbfrom0withk}, the horizontal straight line marks the Big Bang of the
$\Lambda$CDM model, given by (\ref{10.10}). The vertical line marks $r = r_{\rm
AH}$.} \label{drawlumdisconeswithk}
\end{figure}

The resulting light cone profile is shown in Fig. \ref{drawlumdisconeswithk},
compared with the light cone of the $\Lambda$CDM model. The two light cones, as
predicted, do not coincide. In particular, the L--T light cone is everywhere
later than the $\Lambda$CDM cone, and the difference in time increases as the
Big Bang is approached. The L--T light cone meets the Big Bang at a larger value
of $r$, which was seen already in Fig. \ref{drawznointerwithk}. It touches the
$t_B(r)$ curve horizontally, as it should (see Sec. \ref{comments}). However,
$t_B(r)$ asymptotes to a later value of $t$ than the $\Lambda$CDM Big Bang,
namely to $t = - 0.139$ NTU. In consequence, up to a certain $r > r_{\rm AH}$,
the L--T Universe is everywhere younger than $\Lambda$CDM. The difference in the
bang times at the edge of the figure is $\Delta t \approx 0.002$ NTU $\approx
1.96 \times 10^8$ y.

One might suspect that the disagreement between the two light cones is a
consequence of numerical errors. In truth, the numerical error is much smaller,
as shown by the final test: the two sides of eq. (\ref{8.1}) were compared
numerically. The right-hand side, the function $F_r(r) \df {\cal D}/[H_0 (1 +
z)]$, is calculated directly from the input data. The left-hand side, the
function $F_l(r) \df R(t_{\rm ng}(r), r)$, depends on the whole chain of
numerical calculations that were carried out to find $t(r)$. The $F_r(r)$ was
calculated on top of $z_{\rm back}(r)$, see Sec. \ref{numericszwithk}. The
$F_l(r)$ was calculated on top of $t(r)$ and $t_B(r)$. Each time when $t$ and
$t_B$ were found for a given $r$, the corresponding $\eta(r)$ in (\ref{2.7}) was
found by the bisection method (with the precision $\Delta \eta = 10^{-15}$), and
then the $R(t(r), r)$ was calculated from the first of (\ref{2.7}).

\begin{figure}[h]
 ${}$ \\[2cm]
\hspace{-3.5cm}
\includegraphics[scale=0.9]{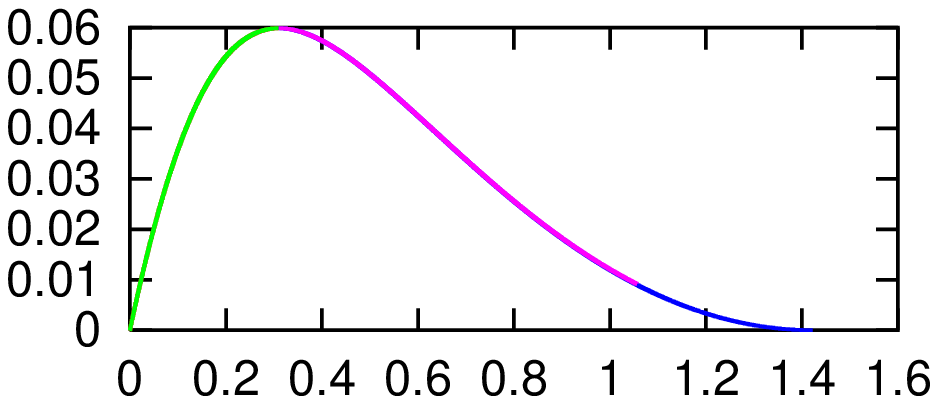}
 ${}$ \\[0.1cm]
\includegraphics[scale=0.4]{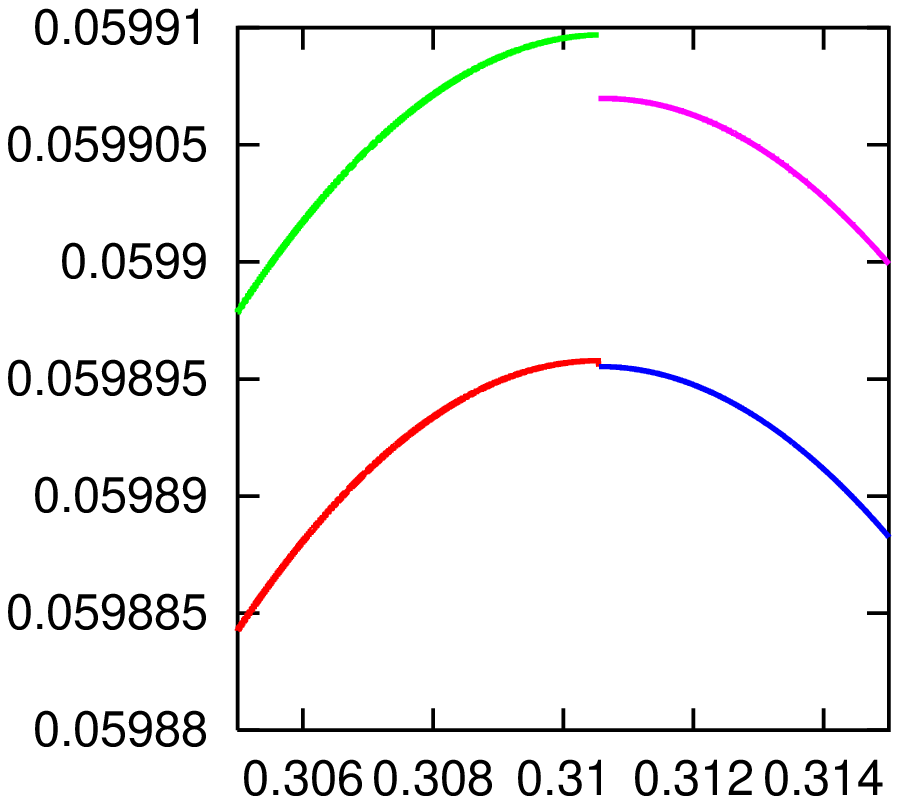}
\includegraphics[scale=0.4]{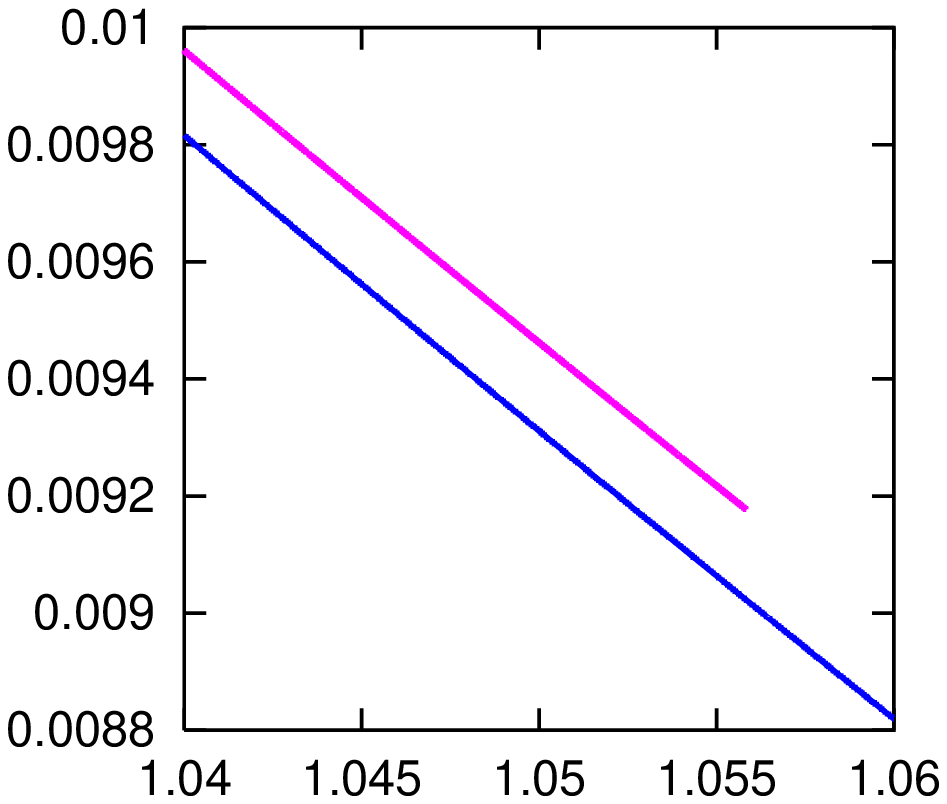}
\caption{{\bf Upper panel:} Comparison of the two sides of eq. (\ref{8.1}). At
this scale, the two functions seem to coincide. {\bf Lower left panel:} closeup
view of the maximum in the upper panel. The upper curve is $F_l(r)$, the lower
curve is $F_r(r)$. The graph shows the discontinuities at the maximum that
resulted from numerical errors. {\bf Lower right panel:} the functions $F_l(r)$
(upper line) and $F_r(r)$ (lower line) at the right end of $F_l$. See text for
more explanation.} \label{preccompare}
\end{figure}

Figure \ref{preccompare} shows the comparison. The upper panel shows the two
functions in full range; at this scale they seem to coincide, except that
$R_l(r)$ ends at a smaller $r$. The lower left panel shows numerical errors at
the maximum. There is a discontinuity in $F_l(r) = R(t(r), r)$ equal to $\approx
2.78 \times 10^{-6}$ NLU $\approx 83.5$ kpc $\approx 2.72 \times 10^5$ y, a
discontinuity in $F_r(r)$, equal to $10^{-7}$ NLU $\approx 3$ kpc $\approx 9800$
y, and the difference between $R_l(r)$ and $R_r(r)$, which, at the maximum, is
$\approx 1.39 \times 10^{-5}$ NLU $\approx 417$ kpc $\approx 1.36 \times 10^6$
y. The lower right panel shows the same difference at the right end of the graph
of $R_l(r)$, which is $\approx 1.44 \times 10^{-4}$ NLU $\approx 4.32$ Mpc
$\approx 1.41 \times 10^7$ y.

In summary: we required that the L--T model with $\Lambda = 0$, $2E/r^2 = - k =$
constant and variable $t_B(r)$ duplicates the $D_L(z)$ function given by
(\ref{2.20}) via (\ref{2.14}). Under these assumptions, the value of ${\cal H}_0
= 67.1$ km/(s $\times$ Mpc) taken from observations \cite{Plan2013} and the
value of $z$ at the AH, calculated from (\ref{9.3}), determine the value of $k$,
and then the shape of $t_B(r)$ is determined such that it mimics the effect of
$\Lambda$ on a single light cone.

Thus, using the $\Lambda = 0$ L--T model with constant $E/r^2 > 0$, one can
explain away the accelerated expansion of the Universe as follows. In the
$\Lambda$CDM model, the bang time is constant, while in the L--T model the Big
Bang occurs progressively later when the position of the observer is approached.
Consequently, the time between the Big Bang and the instant of crossing the
observer's past light cone becomes progressively shorter in L--T than in
$\Lambda$CDM. Because of this, the expansion velocity of matter in the L--T
model, at the points of intersection with this cone, is everywhere greater than
in a Friedmann model with $\Lambda = 0 = k$, and the difference is increasing
toward the observer, similarly to what happens in Fig. \ref{linecompare}. Thus,
accelerating expansion is mimicked: instead of increasing with time, the excess
expansion velocity increases with position in space.

This may look artificial (the observer being placed at that $r$, where $t_B(r)$
is greatest). But it should be noted that the model that led to this conclusion
had from the beginning a built-in artificial assumption, made in order to
simplify the calculations: the function $E$ being the same as in the Friedmann
model. This left the whole task of imitating acceleration to $t_B$ alone. See
also the comments in the next section.

\section{Comments on applications of the L--T model to
cosmology}\label{LTcosmology}

\setcounter{equation}{0}

There is a group of astrophysicists who treat the L--T model as an enemy to kill
rather than a useful device for cosmology. (Example: a quotation from Ref.
\cite{QABC2012}: [the Gaia or E-ELT observatories could distinguish the RW models
from L--T] ``possibly eliminating an {\em exotic alternative explanation to dark
energy}''.) They try to discredit this model in several ways. One of the legends
spread by them says that a realistic L--T model must have constant $t_B$. This
crippling limitation allegedly must be made because $d {t_B} / dr \neq 0$
generates decreasing density perturbations, and they would imply ``extreme''
inhomogeneity at early times \cite{Zibi2011}. This, the argument goes, would
contradict the predictions of inflation.

It is questionable whether the increasing and decreasing density perturbations
can be treated as algebraically independent, and their consequences separately
investigated. This would be correct in a linear theory. In relativity, these two
classes of perturbations are in general present simultaneously, and they
interact nonlinearly. It was proven in Ref. \cite{KrHe2004b} that a non-constant
$t_B$, together with an inhomogeneous $E(r)$, can generate a galaxy cluster out
of a localised, small in amplitude, density or velocity inhomogeneity at the
time of last scattering, without causing any contradiction with the observations
of CMB. The required difference in $t_B$ between the center and the edge of the
cluster is typically below 100 years (in some cases as little as 15 years). So,
clearly, this is not an ``extreme'' inhomogeneity, and at least some processes
taking place after last scattering are compatible with a non-constant $t_B$. The
relevant astrophysical quantity is the density (or velocity) perturbation at the
time of last scattering, and not the gradient of $t_B(r)$.

Formally, the decreasing density perturbation becomes infinite as $t \to t_B$
\cite{PlKr2006}. However, this would be a problem for cosmology if the L--T
model would be supposed to apply all the way to $t = t_B$, which is not the
case. The direct connection between the decreasing/increasing density
perturbation and $\dril {t_B} r \neq 0$ was demonstrated only for the L--T and
Szekeres \cite{Szek1975} models (by Silk \cite{Silk1977} and Goode and
Wainwright \cite{GoWa1982}, respectively, see Ref. \cite {PlKr2006} for short
descriptions). In a more general model, still unknown, which should apply before
recombination, the corresponding connection may be indirect, and need not imply
infinite perturbations close to the Big Bang. And, let us remember, the Big Bang
itself is supposed to go away when quantum gravity provides the right
description of that epoch.

Another widespread belief is that an L--T model mimicking accelerated expansion
contains a void around its center of symmetry; several authors just reflexively
call it a ``void model''. Ref. \cite{ClRe2011} is an example, a few more examples
are listed in Ref. \cite{BCKr2011}. Our result (\ref{13.11}), from which it
follows that $\lim_{r \to 0} \dril {t_B} r < 0$, provides a counterexample to
this belief. Namely, from (\ref{2.8}), knowing that $\lim_{r \to 0} M = \lim_{r
\to 0} R = 0$, we find
\begin{equation}\label{18.1}
\lim_{r \to 0} \frac {R^3} M = \lim_{r \to 0} \frac {3 R^2 R,_r} {M,_r} = \frac
6 {\kappa \rho(t,0)} \df \frac 6 {\kappa \overline{\rho}}.
\end{equation}
Then, using (\ref{2.9}), (\ref{11.4}) and (\ref{18.1}) in (\ref{2.8}), we obtain
\begin{equation}\label{18.2}
\lim_{r \to 0} \left(\kappa \rho,_r\right) = \frac {4 (\kappa
\overline{\rho})^{4/3}} {(6M_0)^{1/3}}\ \sqrt{2M_0 \left(\frac {\kappa
\overline{\rho}} {6 M_0}\right)^{1/3} - k}\ \dr {t_B} r < 0.
\end{equation}
Hence, in this case there is a peak of density at $r = 0$.

Much effort has been spent in the literature on the attempts to disprove the
L--T metric as a viable cosmological model by exploiting its spherical symmetry
(see, for example, Ref. \cite{BCFe2012}). Consequently, it has to be reminded
that this model is mainly used as an exercise to gain insight into a nontrivial
geometry. This insight is then exploited in applying, for example, the Szekeres
model \cite{Szek1975, PlKr2006} to cosmological problems; see examples of such
applications in Refs. \cite{Bole2006} -- \cite{BCMB2013}. The Szekeres model is
a generalisation of L--T; it has no symmetry and is more complicated
computationally. Therefore, insights gained from carrying out the L--T exercises
are helpful.

One more argument against taking literally the predictions of the L--T models
and comparing them with observations interpreted on an FLRW background is given
in Ref. \cite{ClRe2011}. These authors point out that the inclusion of radiation
in the dynamics of spherically symmetric models might upset the results obtained
with radiation neglected. They emphasise that observables deduced from the CMB
have to be recalculated from scratch, and cannot simply be inferred from the
FLRW case.

More generally, the L--T and Szekeres models are not supposed to be \textit{the}
ultimate models of the whole Universe. They are to be understood as {\em{exact
perturbations}} superimposed on the background Friedmann model, and can be
sensibly applied only to the description of local structures, such as galaxy
clusters or voids, see Refs. \cite{KrHe2004b}, \cite{BKHe2005} and
\cite{PlKr2006,BKHC2010}. Consequently, in situations, in which perturbed
Friedmann models are deemed adequate, the L--T and Szekeres models, when they
are correctly understood and applied, can only be still more adequate, being
exact solutions of Einstein's equations. If they are to become objects of the
now-so-called ``precision cosmology'', then results of observations should give
information on the shapes of their arbitrary functions. Outright rejection is
not a constructive approach.

We will never know how good or how bad any given model is until we test it at
full generality in as many situations as will be invented by anyone. This will
help in constructing the next generation of still more precise models. Excluding
elements of a model on the basis of a speculative competing hypothesis is not
what serious science used to be about. And it is unethical to use arguments of
this kind to reject papers submitted for publication, as sometimes happens.

The artificial elements of the model considered here (the $D_L(z)$ being
reproduced only on a single past light cone, the cone reaching the Big Bang
where $\dril {t_B} r = 0$) are present because it was designed to mimic the
observations via their projection on the $\Lambda$CDM past light cone. They do
not appear when the L--T model is directly adapted to observations. The point
made in this paper is: using the L--T model, observations can be accounted for
without introducing the dark energy.

The way, in which the $D_L(z)$ function was reproduced here is not the only one
possible. Iguchi et al. \cite{INNa2002} demonstrated that such a reproduction is
also possible with constant $t_B$, and $E(r)$ designed to mimic the effect of
$\Lambda$. This dual approach will be a subject of a similar analysis as done
here in a future paper.

\appendix

\section{Proof that (\ref{3.2}) has a solution for every
$\Lambda$}\label{AHforeveryLambda}

\setcounter{equation}{0}

Consider the equation equivalent to (\ref{3.2})
\begin{equation}\label{a.1}
F(R) \df \tfrac 1 3 \Lambda R^3 + R - 2M = 0.
\end{equation}
For $\Lambda = 0$ the solution $R = 2M$ obviously exists. In all models with
$\Lambda > 0$ $R$ is oscillating between $R = 0$ and a finite maximal value $R =
R_m$.\footnote{Recall: (\ref{2.2}) has the same algebraic form for the L--T and
Friedmann models. The proof that all models with $\Lambda > 0$ are oscillating
had been given by Friedmann \cite{Frie1922}, see Ref. \cite{PlKr2006}
(Friedmann's cosmological constant $\lambda$ is related to our $\Lambda$ by
$\lambda = - \Lambda$).} At $R = R_m$, where $R,_t = 0$ in (\ref{2.2}), the
following holds
\begin{equation}\label{a.2}
G(R_m) \df \tfrac 1 3 \Lambda {R_m}^3 - 2 ER_m - 2M = 0,
\end{equation}
and there is only one value of $R_m > 0$ that obeys (\ref{a.2}). Thus, at $R =
R_m$
\begin{equation}\label{a.3}
F(R_m) = (2E + 1) R_m.
\end{equation}
Since $2E + 1 \geq 0$ (see (\ref{2.3})), we have $F(R_m) \geq 0$, while at $R =
0$, $F(R) = - 2M \leq 0$ ($F = 0$ only at the center, where $M = 0$). So, $F(R)
= 0$ has a solution for every $\Lambda > 0$; the solution is $R = R_m$ where $E
= -1/2$ and $R < R_m$ where $E > -1/2$. Consequently, an AH exists.

Now consider $\Lambda < 0$. For $0 > \Lambda > \Lambda_{\cal E} \df -
8E^3/(9M^2)$ (the Einstein value, see Ref. \cite{PlKr2006}) the reasoning above
still applies to the oscillating models. For non-oscillating models in the same
range of $\Lambda$, the subcases $E < 0$ and $E \geq 0$ have to be considered
separately. When $E < 0$, the value of $R$ is always greater than the $R_m$
given by (\ref{a.2}), so it follows that $F(R) = 0$ has no solutions in that
range of $R$ (but it had a solution in the range of oscillating models, so the
statement being proven is not contradicted). When $E \geq 0$, $R$ changes
between 0 and $\infty$, and the reasoning given below applies.

For $\Lambda = \Lambda_{\cal E}$ the situation is similar, except that there
exists in addition the static Einstein model, but this does not contradict the
statement being proven.

For $\Lambda < \Lambda_{\cal E}$, $R$ necessarily varies between 0 and $\infty$.
Then, from (\ref{a.1}), $F(0) = -2M \leq 0$, and
\begin{equation}\label{a.4}
\dril F R = \Lambda R^2 + 1.
\end{equation}
This is zero at $R = \pm 1 / \sqrt{- \Lambda}$, so $F$ has a maximum at $R = 1
/\sqrt{- \Lambda}$, and $F(1 /\sqrt{- \Lambda}) = 4 \sqrt{- \Lambda} - 2M$. This
is positive in some range $M \in [0, \sqrt{- \Lambda})$, so $F(R) = 0$ has a
solution in this range, i.e. an AH exists. $\square$

\section{Proof that the AH coincides with the set $R,_{tr} = 0$ only in
exceptional cases}\label{whereAH}

\setcounter{equation}{0}

Calculating $R,_{tr}$ from (\ref{11.3}) and taking the result at $R = 2M$, one
obtains, using (\ref{2.2}) with $\Lambda = 0$
\begin{eqnarray}\label{b.1}
&& \left.R,_{tr}\right|_{\rm AH} = \left\{\frac {E,_r} {2E} \sqrt{2E + 1}\right.
\\
&& + \left. \frac 1 {4M} \left[\left(\frac 3 2 \frac {E,_r} E - \frac {M,_r}
M\right) \left(t - t_B\right) - t_{B,r}\right]\right\}_{\rm AH}. \nonumber
\end{eqnarray}
For $E = 0$, the analogue of (\ref{11.3}) is (\cite{PlKr2006}, eq. (18.112))
\begin{equation}\label{b.2}
R,_r = \frac {M,_r} {3M}\ R - \sqrt{\frac {2M} R} t_{B,r}.
\end{equation}
The cases $E > 0$, $E = 0$ and $E > 0$ must be considered separately. Only the
case $E < 0$ is presented here; the corresponding result for $E > 0$ follows
analogously, and the one for $E = 0$ follows easily from (\ref{b.2}).

For $E < 0$ one finds $(t - t_B)$ as a function of $R$ from (\ref{2.5}) and
takes it at $R = 2M$, obtaining
\begin{eqnarray}\label{b.3}
&& \left(t - t_B\right)_{\text{AH}} = \frac M {(- 2E)^{3/2}} \nonumber \\
&& \ \ \ \ \ \times \left[\arccos\left(1 + 4E\right) - 2 \sqrt{- 2E (1 +
2E)}\right]
\end{eqnarray}
(the $\arccos$ is to be calculated for $0 \leq 1 + 2ER/M \leq \pi$, i.e. for the
expanding phase of the Universe). After substituting this in (\ref{b.1}) the
following is obtained
\begin{eqnarray}\label{b.4}
&& \left.R,_{tr}\right|_{\rm AH} = \frac {t_{B,r}} {4M}  \nonumber \\
&&\ \ \  - \frac 1 {4 (-2E)^{3/2}} \left(\frac 3 2 \frac {E,_r} E - \frac {M,_r}
M\right) \arccos\left(1 + 4E\right)  \nonumber \\
&&\ \ \  - \frac 1 {4E} \left(\frac {E,_r} {2 E} - \frac {M,_r} M\right)
\sqrt{2E + 1}.
\end{eqnarray}
The AH is a curve in the $(t, r)$ plane, so if $R,_{tr} = 0$ should hold along
the whole AH, (\ref{b.4}) would force a relation between $M$, $E$ and $t_B$,
thus reducing the number of arbitrary functions to 2. This means that $R,_{tr} =
0$ can hold along the AH only in special cases.

The corresponding equation for $E = 0$ is
\begin{equation}\label{b.5}
t_{B,r} = 2 M,_r / 3,
\end{equation}
from (\ref{b.2}), and it also limits the generality of the model.

Note that (\ref{b.4}) and (\ref{b.5}) do not hold in the Friedmann model, where
$M/r^3 = M_0$, $2E/r^2 = -k$ and $t_B$ are constant. Thus,
$\left.R,_{tr}\right|_{\rm AH} \neq 0$ even in the Friedmann limit.

\section{Proof that (\ref{13.5}) has only one real solution $X > 0$}\label{oneX}

\setcounter{equation}{0}

Let us write (\ref{13.3}) as
\begin{equation}\label{c.1}
f(X) \df X^3 + kX - b = 0,
\end{equation}
where $k < 0$ and $b \df 2 M_0 H_0 > 0$. The function $f(X)$ has a local maximum
at $X_- = - \sqrt{- k /3} < 0$ and a local minimum at $X_+ = \sqrt{- k /3} > 0$.
We have
\begin{equation}\label{c.2}
f(X_+) = - \frac {2 (- k)^{3/2}} {3 \sqrt{3}} - b < 0,
\end{equation}
so there must be a zero of $f(X)$ in $(X_+, + \infty)$, and
\begin{equation}\label{c.3}
f(X_-) = \frac {2 (- k)^{3/2}} {3 \sqrt{3}} - b.
\end{equation}
When $(f(X_-) < 0$, there are no more real zeros of $f(X)$. When $(f(X_-) = 0$,
$X = X_-$ is a double real zero of $f(X)$, additional to that guaranteed by
(\ref{c.2}). When $(f(X_-)> 0$, there are two more real zeros of $f(X)$.
However, the additional zeros are at $X < 0$, since $f(0) < 0$. $\square$

\section{The derivation of (\ref{13.11})}\label{limof1311}

\setcounter{equation}{0}

Using (\ref{13.1}) and (\ref{13.5}) we find
\begin{equation}\label{d.1}
\lim_{r \to 0} \sqrt{\frac {2M_0 H_0 r (1 + z)} {\cal D} - k} = X.
\end{equation}
Using (\ref{13.7}), (\ref{11.10}), (\ref{9.5}) and (\ref{d.1}) in (\ref{11.14})
we obtain
\begin{eqnarray}\label{d.2}
&& \lim_{r \to 0} \dr {t_B} r = \frac {2M_0} {-3M_0 H_0 + kX} \lim_{r \to 0}
\left\{\frac 1 r \left[{\cal D} B_3 - \frac 3 2\ A_1 \right.\right. \nonumber \\
&& + \left.\left. \frac {kr^2 A_1} {\left(A_2 + 1\right)^2}\right]\right\} \df
\frac {2M_0} {-3M_0 H_0 + kX}\ {\cal Z}.
\end{eqnarray}
In what follows, two more new symbols will be used:
\begin{eqnarray}
Q &\df& \frac 1 {\sqrt{\Omega_m (1 + z)^3 + \Omega_{\Lambda}}}, \label{d.3} \\
{\cal U} &\df& \sqrt{\frac {2M_0 H_0 r (1 + z)} {\cal D} - k}. \label{d.4}
\end{eqnarray}
After writing out $B_3$, $A_1$ and $A_2$ we find from (\ref{d.2})
\begin{eqnarray}\label{d.5}
{\cal Z} &=& \lim_{r \to 0} \left\{\frac 1 r\ \left[(1 + z) Q - \frac {k {\cal
D} Q} {2M_0 H_0 r}\right.\right. \nonumber \\
&-& \left.\left.\frac {{\cal D}^2 \sqrt{1 - kr^2} {\cal U}} {2M_0 H_0 r^2 (1 +
z)}\right]\right\}.
\end{eqnarray}
The limit at $r \to 0$ of the expression in square brackets is zero, so we can
apply the de l'H\^{o}pital rule and obtain
 \begin{widetext}
\begin{eqnarray}\label{d.6}
{\cal Z} &=& \lim_{r \to 0} \left\{\dr z r\ \left[Q - \frac 3 2 \Omega_m (1 +
z)^3 Q^3 +\frac {{\cal D}^2 \sqrt{1  - kr^2} {\cal U}} {2M_0 H_0 r^2 (1 + z)^2}
- \frac {{\cal D} \sqrt{1 - kr^2}} {2 r (1 + z) {\cal U}} + \frac 3 4\ k
\Omega_m \frac {(1 + z)^2 {\cal D} Q^3} {M_0 H_0 r}\right] \right. \nonumber \\
&+& \left. \frac {k
{\cal D}^2 {\cal U}} {2M_0 H_0 r (1 + z) \sqrt{1 - k r^2}}\right\} \nonumber \\
&+& \lim_{r \to 0} \left[\frac {kQ} {2 M_0 H_0} - \frac {k {\cal D} \sqrt{1 - k
r^2}} {M_0 H_0 r (1 + z) {\cal U}} + \frac {3 \sqrt{1 - k r^2}} {2 {\cal
U}}\right] \times \lim_{r \to 0} \left[\frac 1 r\ \left(\frac {\cal D} r - Q \dr
z r\right)\right],
\end{eqnarray}
 \end{widetext}
where the expression in the first two lines and the first limit in the third
line can be readily calculated:
\begin{equation}\label{d.7}
{\cal Z} = F_1 + \left(\frac 3 {2X} - \frac k {2M_0 H_0}\right) \lim_{r \to 0}
\left[\frac 1 r\ \left(\frac {\cal D} r - Q \dr z r\right)\right],
\end{equation}
where
\begin{equation}\label{d.8}
F_1 \df X \left[\frac 3 2 \left(1 - \Omega_m\right) - \frac {kX} {2 M_0 H_0}
\left(1 - \frac 3 2 \Omega_m\right)\right].
\end{equation}
In (\ref{d.7}) we now substitute for $\dril z r$ from (\ref{11.12}), then factor
out $1 / (rB_3)$ and use (\ref{13.7}). The result is
\begin{equation}\label{d.9}
{\cal Z} = F_1 - \lim_{r \to 0} \left\{\frac {{\cal D} B_3} r - \frac {Q B_2 (1
+ z)} {r \sqrt{1 - k r^2}} \left[\frac 3 2 - \frac {kr^2} {\left(A_2 +
1\right)^2}\right]\right\}.
\end{equation}
Comparing (\ref{d.9}) with (\ref{d.2}) we see that
\begin{eqnarray}\label{d.10}
&& {\cal Z} = F_1 - {\cal Z} \\
&& - \lim_{r \to 0} \left\{\frac 1 r \left[\frac 3 2 - \frac {kr^2} {\left(A_2 +
1\right)^2}\right] \left[A_1 - \frac {Q B_2 (1 + z)} {\sqrt{1 - k
r^2}}\right]\right\}. \nonumber
\end{eqnarray}
The second factor in square brackets has the limit zero, the first one is
finite. Consequently
\begin{equation}\label{d.11}
{\cal Z} = \frac 1 2\ F_1 = \frac 1 4\ X \left[3 \left(1 - \Omega_m\right) -
\frac {kX} {M_0 H_0} \left(1 - \frac 3 2 \Omega_m\right)\right].
\end{equation}
Using this in (\ref{d.2}) we obtain (\ref{13.11}). $\square$

 \bigskip

{\bf Acknowledgement.}  I thank Krzysztof Bolejko for several useful comments.

\end{document}